\documentclass[12pt]{article}
\pdfoutput=1
\usepackage{amsmath,amssymb,epsfig,amsfonts}
\usepackage{graphicx,subfigure}
\usepackage[usenames, dvipsnames]{color}
\usepackage[backref]{hyperref}
\usepackage{cite}
\usepackage{verbatim} 
\usepackage{float}
\restylefloat{table}
\usepackage[all]{xy}
\usepackage{pdflscape}
\usepackage{tikz}
\usepackage{array}
\usepackage{youngtab}
\usepackage{multirow}
\usepackage{color}
\usepackage{ulem}

\def\Sone{\mathbb{S}^1}

\newcommand{\rem}[1]{} 

\def\Z{\mathbb{Z}}

\def\R{\mathbb{R}}
\def\N{\mathcal{N}}
\def\P{\mathbb{P}}

\def\rk{\operatorname{rk}}

\def\Hirz[#1]{\mathbbm{F}_{#1}}
\def\o[#1]{\overline{#1}}

\def\coker{\mbox{coker}}

\def\ker{\mbox{ker}}

\newcommand{\be}{\begin{equation}}
\newcommand{\ee}{\end{equation}}
\newcommand{\ba}{\begin{aligned}}
\newcommand{\ea}{\end{aligned}}

\newcommand{\bea}{\begin{eqnarray}}
\newcommand{\eea}{\end{eqnarray}}

\makeatletter
\newcommand\xleftrightarrow[2][]{%
  \ext@arrow 9999{\longleftrightarrowfill@}{#1}{#2}}
\newcommand\longleftrightarrowfill@{%
  \arrowfill@\leftarrow\relbar\rightarrow}
\makeatother

\makeatother

\begin{document}

\baselineskip=18pt  
\numberwithin{equation}{section}  


\thispagestyle{empty}
\vspace*{0.8cm} 
\begin{center}
 {\LARGE 
Compact, Singular $G_2$-Holonomy Manifolds and \\
\bigskip
M/Heterotic/F-Theory Duality}

 \vspace*{1.8cm}
{Andreas P. Braun$\,^1$  and Sakura Sch\"afer-Nameki$\,^2$}\\

 \vspace*{1.2cm} 
{\it $^1$ 
Rudolf Peierls Centre for Theoretical Physics,\\ 
University of Oxford, Oxford, OX1 3NP, UK}\\
 {\tt andreas.braun\phantom{@}physics.ox.ac.uk}\\
  
\bigskip
{\it $^2$ Mathematical Institute, University of Oxford \\
 Woodstock Road, Oxford, OX2 6GG, UK}\\
  {\tt {gmail:$\,$sakura.schafer.nameki}}\\
\vspace*{0.8cm}
\end{center}
\vspace*{.5cm}
%
\noindent 
We study the duality between M-theory on compact holonomy $G_2$-manifolds and the heterotic string on Calabi-Yau three-folds. 
The duality is studied for K3-fibered $G_2$-manifolds, called twisted connected sums, which lend themselves to an application of fiber-wise M-theory/Heterotic Duality. For a large class of such $G_2$-manifolds we are able to identify the dual heterotic as well as F-theory realizations. 
First we establish this chain of dualities for smooth $G_2$-manifolds. This has a natural generalization to situations with non-abelian gauge groups, which correspond to singular $G_2$-manifolds, where each of the K3-fibers degenerates. We argue for their existence through the chain of dualities, supported by non-trivial checks of the spectra. The corresponding 4d gauge groups can be both Higgsable and non-Higgsable, and we provide several explicit examples of the general construction. 


\newpage

\tableofcontents
\section{Introduction}

Exploring and developing tools to characterize minimally supersymmetric theories, in particular super-conformal theories (SCFTs), is a challenging task. The successes of studying $\N\geq 2$ supersymmetric theories, by realizing them geometrically within string theory motivates an approach, that gives rise to a similarly geometric chacterization of the moduli spaces of $\N=1$ theories. Here we would like to focus on 4d $\N=1$ theories. A powerful approach to studying and classifying 6d $\N=(1,0)$ SCFTs is provided by F-theory \cite{Heckman:2013pva}, which heavily relies on the purely geometric nature of the classification problem, complemented with anomaly considerations. In 4d F-theory vacua, the geometry is only one part of the story, and needs to be supplemented with four-form $G_4$-flux data. This somewhat complicates the F-theoretic approach. 

An alternative way, which is largely geometric, and yields 4d $\N=1$ vacua, is M-theory on 7-manifolds with $G_2$-holonomy. Here the geometry of the manifold determines both the gauge group (from codimension 6 singularities), and if present, the spectrum of chiral matter (codimension 7 singularities), without the necessity of turning on fluxes \cite{Acharya:2001gy}. The shortcomings of this approach are thus far the absence of any compact models, which have singularities to engineer chirality in 4d. String compactifications on $G_2$-manifolds result in either 3d $\N=1$ (for heterotic) or 3d $\N=2$ (for Type II strings) theories \cite{Shatashvili:1994zw}, whose moduli space would indeed also be of great interest to explore further. 
For a review of the works until the early 2000s we refer to \cite{Acharya:2004qe}. 

Recently progress in the construction of compact $G_2$-manifolds has been made within mathematics \cite{Corti:2012kd, MR3109862}, based on earlier results by Kovalev \cite{MR2024648}. These so-called twisted connected sum (TCS) $G_2$-manifolds have a K3-fiberation over a three-manifold. In the physics literature, M-theory compactifications on such geometries were studied in \cite{Halverson:2014tya, Halverson:2015vta, Guio:2017zfn} and in string theory compactifications \cite{Braun:2017ryx}. 

The K3-fibration of the TCS geometries is particularly suggestive within the context of M-theory compactifications, as it is amenable to M-theory/heterotic duality. This duality states that M-theory on a K3-surface has the same low energy effective theory as the heterotic string on $T^3$ \cite{Witten:1995ex}. Applying this duality fiberwise over a three-dimensional base, suggests the duality between M-theory on a TCS $G_2$-manifold and heterotic string theory on a Calabi-Yau manifold, which has a Strominger-Yau-Zaslow (SYZ) fibration \cite{Strominger:1996it}, i.e. a special Lagrangian $T^3$-fibration over a three-dimensional base:
 \be
\hbox{M-theory on TCS $G_2$-manifold $\quad \longleftrightarrow\quad$ Heterotic on SYZ-fibered Calabi-Yau three-fold}\,.
 \ee
This is the duality that we will explore in the present paper. 

The main result of our work is the construction of compact TCS $G_2$-manifolds, along with their heterotic and F-Theory duals. Our identification of dual geometries is substantiated by a general proof that the spectra of the dual compactifications agree. In particular, we are able to extend this duality to non-abelian theories, i.e. TCS $G_2$-manifolds, which have singular K3-fibers over the entire three-manifold base. This does not yet realize the ever so elusive chiral fermions from a compact geometry, however it provides a construction of a singular $G_2$, which has non-abelian gauge groups. The goal of this work is to setup the M-theory/Heterotic duality in the context of the TCS construction including singular K3-fibers, thus setting the stage for potentially modifying the geometry to include conical singularities such as proposed in the local modes of \cite{Witten:2001uq, Acharya:2001gy}. 

Among the non-abelian gauge theories we engineer, there are both Higgsable and non-Higgsable models. From an F-theoretic point of view, the non-Higgsable models constructed here are examples of geometrically non-Higgsable clusters \cite{Grassi:2014zxa, Morrison:2014lca} for Calabi-Yau fourfolds. In addition, we show that in the present case these theories are also non-Higgsable in the 4d effective theory. 


The plan of this paper is as follows. Section \ref{sec:Background} reviews the TCS construction of $G_2$ holonomy manifolds, as well as the duality between M-theory on K3 and heterotic on $T^3$, and outlines the fiber-wise duality which is applicable when the K3s in the M-theory geometry are elliptically fibered. In section \ref{sec:TCSSYZ} we then construct the heterotic dual to a smooth TCS $G_2$-manifold, where the geometry is the Schoen Calabi-Yau three-fold. Using the duality to heterotic we then construct TCS fibrations with singular K3-fibers in section \ref{sec:NHC}, which have non-abelian gauge groups in 4d, that we show to be non-Higgsable. To solidify the study of these singular K3-fibered $G_2$ compacitifications, we also determine the F-theory dual for each of these models in section \ref{sect:F-theory}. 
In section \ref{sec:GeneralStory} this construction is generalized to M-theory on TCS $G_2$s whose building block realize arbitrary not-necessarily non-Higgsable gauge groups. 
This setup provides a multitude of generalizations and extensions, some of which are discussed in section \ref{sec:Disc}. 


\section{M/Het Duality and TCS $G_2$-manifolds}
\label{sec:Background}

We study M-theory on manifolds with $G_2$ holonomy, which are constructed as twisted connected sums with K3-fibered building blocks. We map this by fiber-wise application of the 7d heterotic/M-theory duality to the heterotic string on an SYZ-fibered Calabi-Yau three-fold. In this section we review the TCS construction for $G_2$-manifolds, as well as the standard heterotic on $T^3$ to M-theory on K3 duality. 


\subsection{TCS $G_2$-Manifolds}
\label{sec:TCSG2}

To begin with, we review the basics of the twisted connected sum (TCS) construction of $G_2$ holonomy manifolds 
following \cite{Corti:2012kd, MR3109862, MR2024648}. Let 
\be\label{Zpm}
Z_{\pm} = K3\stackrel{\pi_\pm}{\longrightarrow}  \mathbb{P}^1
\ee
be two holomorphic three-folds which are smooth and K3-fibered over $\mathbb{P}^1$, and have first Chern class 
\be\label{c1Z}
c_1 (Z_{\pm}) = [S_\pm]\,,
\ee
where $[S_\pm]$ is the class of the generic K3-fiber. The $Z_\pm$ are called building blocks. Note that due to (\ref{c1Z}) these are not Ricci-flat, as the K3-fiber is only twisted half as much as needed for a Calabi-Yau three-fold. Furthermore, we require that $H^3(Z_\pm,\mathbb{Z})$ is torsion free and that the image, $N_\pm$, of 
\be
\rho_{\pm} :\qquad H^2 (Z^\pm, \mathbb{Z}) \ \rightarrow H^2 (S^\pm_0, \mathbb{Z}) \cong \Lambda \equiv U^{\oplus 3} \oplus (-E_8)^{\oplus 2} \,,
\ee
is primitive in $H^2(S_0^\pm,\mathbb{Z})$, i.e. $\Lambda / N_{\pm}$ has no torsion. The orthogonal complement of $N_\pm$ in $\Lambda$ is denoted by $T_{\pm}$.

From this we can construct two asymptotically cylindrical (acyl) Calabi-Yau (CY) three-folds 
\be
X_{\pm}= Z_{\pm} \setminus S_0^{\pm}\,,
\ee 
where $S_{0}^{\pm}$ are smooth fibers above points $p_0^\pm$ in the $\mathbb{P}^1$ bases of $Z_\pm$. A $G_2$-manifold is then obtained as a twisted connected sum of $X_\pm \times \Sone_{e\pm}$ \footnote{We will denote circles by $\Sone$ in the following.} as follows:
in the vicinity of the excised fiber $S_0^\pm$, $X_\pm$  has an asymptotically cylindrical region in which it approaches 
$S_{0\pm} \times \Sone_{b\pm}  \times \mathbb{R}^+$. 
In these asymptotic regions, the two building blocks $X_\pm \times \Sone_{e\pm}$ are identified by exchanging 
$\Sone_{e\pm}$ with the circles $\Sone_{b\pm}$, while the asymptotic K3-fibers are mapped to each other by a 
hyper-K\"ahler rotation, also called a Donaldson matching:
\be\label{eq:DM}
g^*:\qquad \omega_{S_0^{\pm}} \ \longleftrightarrow \ \Re \Omega_{S_0^{\mp}}\,,\qquad 
\Im (\Omega_{S_0^{\pm}})  \ \longleftrightarrow \ - \Im (\Omega_{S_0^{\mp}}) \,.
\ee
Here, $\omega$ is the K\"ahler form, and $\Omega^{(2,0)} = \Re \Omega + i \Im \Omega$ the holomorphic two-form of the respective K3-fiber. 
The resulting 7-manifold $J$ was shown to admit a $G_2$ holonomy metric \cite{Corti:2012kd, MR3109862, MR2024648}. 

The identification between $\Sone_{b\pm}$ and $\Sone_{e\pm}$ results in a three-sphere $S^3$ glued together from two  solid tori. The $G_2$ holonomy manifold then can be viewed as a  K3-fibration over $S^3$
\be\label{K3J}
 K3\   \hookrightarrow \ J\  \rightarrow\  S^3 \,.
\ee
The K3-fibers are conjectured to be co-associative four-cycles in the $G_2$ manifold, i.e. they are calibrated with $\star \Phi_3$, where $\Phi_3$ is the $G_2$ form.

By the Torelli theorem, the diffeomorphism $g$ is induced by a lattice isometry 
\be
 g_{\Lambda}:\qquad H^{2} (S_{0}^+, \mathbb{Z}) \ \rightarrow \ H^2 (S_0^- , \mathbb{Z}) \,,
\ee
and we can think of this lattice isometry as a being in turn induced from a common embedding of $N_\pm$ into $\Lambda$.

Using the Mayer-Vietoris sequence then allows to compute the integral cohomology groups of $J$ in terms of the data of the building blocks and the Donaldson matching as  \cite{Corti:2012kd, MR3109862, BdZ}
\be\label{eq:bettinumbersTCS}
\ba
H^1(J,\mathbb{Z}) & =   0 \\
H^2(J,\mathbb{Z}) & =  N_+ \cap N_- \oplus K(Z_+) \oplus K(Z_-) \\
H^3(J,\mathbb{Z}) & = \mathbb{Z}[S] \oplus \Gamma^{3,19} /(N_+ + N_-) \oplus (N_- \cap T_+) \oplus (N_+ \cap T_-)\\
& \hspace{1cm} \oplus H^3(Z_+)\oplus H^3(Z_-) \oplus K(Z_+) \oplus K(Z_-) \\
H^4(J,\mathbb{Z}) & = H^4(S) \oplus (T_+ \cap T_-) \oplus \ \Gamma^{3,19} /(N_- + T_+) \oplus \Gamma^{3,19} /(N_+ + T_-) \\
& \hspace{1cm} \oplus  H^3(Z_+)\oplus H^3(Z_-) \oplus K(Z_+)^* \oplus K(Z_-)^* \\
H^5(J,\mathbb{Z}) & = \Gamma^{3,19} /(T_+ + T_-) \oplus K(Z_+) \oplus K(Z_-) \,.
\ea
\ee
Here 
\be
K(Z_\pm)= 
\ker(\rho_\pm)/[S_\pm] \,,
\ee
is the lattice, which is generated by the number of homologically independent components of reducible K3-fibers in $Z_\pm$, modulo the class of the fibre. For orthogonal matchings, where
\begin{equation}\label{eq:orthgluecond}
N_\pm \otimes \R = (N_\pm\otimes\R)\cap(N_\mp\otimes\R) \oplus (N_\pm\otimes\R \cap T_\mp\otimes\R) \, ,
\end{equation}
the two independent Betti numbers obey the simple equations:
\begin{equation}\label{eq:b2b3sumruleorthgluing}
\begin{aligned}
 b_2 & = |K(Z_+)| + |K(Z_-)| + |N_+\cap N_-| \\
 b_3 & = 23 + 2(h^{2,1}_+ +h^{2,1}_-) + (|K(Z_+)| + |K(Z_-)|) -  |N_+\cap N_-|  \, .
\end{aligned}
\end{equation}
We will only consider such matchings in this paper. 

Building blocks may be realized from blowups of semi-Fano threefolds \cite{MR3109862}, but may also be realized as toric hypersurfaces, which in turn have an elegant combinatorial description \cite{Braun:2016igl}. 
We will make extensive use of this method to construct and analyse building blocks with specific properties, so what we summarize it here for completeness, however for a detailed exposition we refer the reader to \cite{Braun:2016igl}. We will provide detailed examples in the subsequent sections, which should make the method accessible. 

The starting point of the construction is a pair of four-dimensional lattice polytopes $\Diamond^\circ$ and $\Diamond$ in dual lattices $N$ and $M$, respectively, obeying 
\begin{equation}\label{eq:topsduality}
 \begin{aligned}
 & \langle \Diamond, \Diamond^\circ \rangle \geq -1 & \\
  \langle  \Diamond,\nu_e \rangle \geq 0 \hspace{.5cm} & &  \langle m_e, \Diamond^\circ\rangle \geq 0
\end{aligned}
\end{equation}
with the choice $m_e = (0,0,0,1)$ and $\nu_e = (0,0,0,-1)$. Such polytopes are called a dual pair of tops. Similar to the construction of \cite{Batyrev:1994hm}, $\Diamond$ defines a compact but generally singular toric variety through its normal fan $\Sigma_n(\Diamond)$, together with a family of hypersurfaces $Z_s$. Each $k$-dimensional  face $\Theta$ of $\Diamond$ is associated with a $(4-k)$ dimensional cone $\sigma(\Theta)$ of the normal fan $\Sigma_n(\Diamond)$. The resulting variety $Z_s$ can be crepantly resolved into a smooth manifold $Z$ by refining the fan $\Sigma \rightarrow \Sigma_n$ using all lattice points on $\Diamond^\circ$ as rays. Such manifolds have all of the properties of building blocks.
 Concretely, the defining equation of the resolved hypersurface is
\begin{equation}\label{HypSurf}
Z: \,\,\, 0 = \sum_{m \in \Diamond} c_m z_0^{\langle m, \nu_0 \rangle}\prod_{\nu_i \in \Diamond^\circ} z_i^{\langle m, \nu_i\rangle +1} \, .
\end{equation}
Here, $m$ are lattice points on $\Diamond$, $c_m$ are generic complex coefficients, the $z_i$ are the homogeneous coordinates associated with lattice points $\nu_i$ on $\Diamond^\circ$, and $z_0$ is the homogeneous coordinate associated with the ray through $\nu_0 = (0,0,0,-1)$.  
Note that $Z$ as defined in this way has first Chern class given by the fiber class, which equals to $[z_0]$. 
\footnote{In the standard Calabi-Yau hypersurface case the power of $z_0$ would be $\langle m, \nu_0 \rangle +1$. The absence of the $+1$ indicates that the first Chern class of $Z$ is non-trivial and given by $[z_0]$. }

We can compute the Hodge numbers of $Z$ as well the ranks of the lattices $N$ and $K$ in purely combinatorial terms. The result is that $h^{i,0}(Z) =  0$ and \cite{Braun:2016igl,Braun:2017ryx}
\begin{equation}\label{eq:bbtopology}
\begin{aligned}
h^{2,1} (Z)& = \ell(\Diamond) - \ell(\Delta_F) + \sum_{\Theta^{[2]} < \Diamond} \ell^*(\Theta^{[2]})\cdot \ell^*(\sigma_n(\Theta^{[2]})) - \sum_{\Theta^{[3]} < \Diamond}\ell^*(\Theta^{[3]}) \cr 
h^{1,1} (Z) &= -4 + \sum_{\Theta^{[3]} \in \Diamond}  1 
+ \sum_{\Theta^{[2]}\in\Diamond} \ell^* (\sigma_n (\Theta^{[2]})) 
+ \sum_{\Theta^{[1]}\in\Diamond} \ell^* (\sigma_n (\Theta^{[1]}) +1) (\ell^* (\sigma_n(\Theta^{[1]}))) 
\cr  
|N| &=  \ell(\Delta_F^\circ) - \sum_{\Theta_F^{[2]} < \Delta_F^\circ} \ell^*(\Theta_F^{[2]}) - 4 + \sum_{\mbox{ve}\,\, \Theta_{F}^{\circ [1]} < \Delta_F^\circ}  \ell^*(\Theta_F^{[1]})\ell^*(\Theta_F^{\circ [1]}) \cr
|K| &= h^{1,1}(Z) - |N| -1 \cr 
&= \ell(\Diamond^\circ) - \ell(\Delta_F^\circ) + \sum_{\Theta^{\circ [2]} < \Diamond^\circ} \ell^*(\Theta^{\circ [2]})\cdot \ell^*(\sigma_n(\Theta^{\circ [2]})) - \sum_{\Theta^{\circ[3]} < \Diamond^\circ}\ell^*(\Theta^{\circ[3]}) \,.
\end{aligned}
\end{equation}
The quantities in these relations are defined as follows. $\Delta_F$ is the subpolytope of $\Diamond$, which is orthogonal to $\nu_e$. It is reflexive and determines the algebraic family in which the K3 fibres are contained via the construction of \cite{Batyrev:1994hm}. The $k$-dimensional faces of $\Diamond$ are denoted by $\Theta^{[k]}$ and $\sigma_n(\Theta)$ is the cone in the normal fan associated with each $\Theta$. The function $\ell$ counts the number of lattice points on a polytope, while the function $\ell^*$ counts the number of lattice points in the relative interior. In the case of $\ell^*(\sigma_n(\Theta^{[k]}))$ this refers to the points on $\Diamond^\circ$ in the relative interior of $\sigma_n(\Theta^{[k]})$. Finally,
in the expression for $|N|$, the sum only runs over vertically embedded (ve) faces $\Theta^{\circ[1]}$ of $\Delta_F^\circ$, which are those faces bounding a face $\Theta^{\circ[2]}$ of $\Diamond^\circ$ which is perpendicular to $F = \nu_e^\perp$, see \cite{Braun:2016igl,Braun:2017ryx} for a detailed exposition. The faces $\Theta^{\circ[1]}$ and $\Theta^{[1]}$ are the unique pair of faces on $\Delta_F,\Delta_F^\circ$ obeying $\langle \Theta^{[1]},\Theta^{\circ [1]} \rangle$.

Note that exchanging the roles of $\Diamond$ and $\Diamond^\circ$ exchanges $h^{2,1}$ and $|K|$ but keeps $b_2+b_3$ invariant, which is relevant for $G_2$ mirror symmetry \cite{Braun:2017ryx}.

\subsection{M-theory on K3/Heterotic on $T^3$}

Let us review briefly the standard 7d duality with 16 supersymmetries between the heterotic string theory and M-theory. 
For the remainder we will focus on the $E_8\times E_8$ heterotic string for concreteness. 
The first evidence for this duality is the observation that the moduli space of the heterotic string on $T^3$ agrees with moduli space as M-theory on K3 \cite{Witten:1995ex} and is given by
\be
\mathcal{M} = \mathbb{R}^+ \times \mathcal{M}_1 \equiv \mathbb{R}^+ \times \left[SO(3,19, \mathbb{Z}) \left\backslash SO( 3, 19 ,\mathbb{R})  \right/ SO(3, \mathbb{R}) \times SO(19, \mathbb{R})\right] \,.
\ee
Here $\mathbb{R}^+$ corresponds to the heterotic string coupling, and the volume of the K3, respectively, and the remaining part $\mathcal{M}_1$ is the Narain moduli space of the heterotic $T^3$ compactification, i.e. the moduli space of signature $(3, 16+3)$ of momentum and winding modes, modulo the action of the T-duality group $SO(3, 19, \mathbb{Z})$. On the dual side this corresponds to the moduli space of Einstein metrics on K3 with unit volume. Note that Einstein implies automatically Ricci-flat on a K3 \cite{Aspinwall:1996mn}.  

By matching the effective actions, the fields are mapped as follows:
\be
\lambda_{het} = e^{3 \gamma} \,.
\ee
where $\lambda_{het}$ is the heterotic string coupling and $e^{\gamma}$ is the radius of the K3 surface. The other fields are mapped as follows: the 10d heterotic string with $G= E_8 \times E_8$ gauge symmetry, has a metric $g_{\mu\nu}$, anti-symmetric 2-form $B_{\mu\nu}$, dilaton $\phi$ and gauge field $A_\mu$. Upon dimensional reduction on $T^3$, the 7d fields are ($y^i, \ i = 0, \cdots, 6$ label the coordinates on $\mathbb{R}^7$ and $\hat{\alpha}=1, \cdots, 3$ with $e^{\hat{\alpha}}$ the drei-beine and $\varpi^{(k)}$ denotes the harmonic $k$-forms along the torus)
\be
\ba
g_{\mu\nu} dx^\mu dx^\nu &=  g_{ij} dy^i dy^j  + \sum_{\hat{\alpha}=1}^3 v_i^{\hat{\alpha}} dy^i e_{\hat{\alpha}} +  \sum_{\hat{\alpha},\hat{\beta}=1}^3\varphi^{\hat{\alpha}\hat{\beta}} e_{\hat{\alpha}} e_{\hat{\beta}} \cr  
B_{\mu\nu} dx^\mu \wedge dx^\nu &= b_{ij} dy^i \wedge dy^j + \sum_{\alpha= 1}^3 b_{i}^{\alpha}\wedge \varpi_\alpha^{(1)} + \sum_{a=1}^3\beta^a \varpi_{a}^{(2)} \cr 
A_\mu dx^\mu &=  A_i dy^i +  \sum_{a=1}^3 \sum_{I=1}^{16} a_I^a  t^I_a\cr 
e^{\phi}& = \lambda_{het}\,.
\ea
\ee
The scalar fields are the dilaton, six scalars from the metric component $\varphi^{\hat{\alpha}\hat{\beta}}$ along the torus, and three $\beta$ from the B-field, expanded along the harmonic $2$-forms on the $T^3$. The gauge field gives rise to $a_I^a$, which are the  three Wilson lines along the Cartan subalgebra $\mathfrak{u}(1)^{16}$ of the gauge group. Overall, the dimension is 58, which agrees with that of $\mathcal{M}$. This does not yet prove the coset structure, which follows by careful analysis of the points of symmetry enhancement in the moduli space \cite{Narain:1985jj}. 

The gauge fields in 7d are from the unbroken gauge group, determined by the commutant of the Wilson lines inside $G$ together with the gauge fields obtained from the metric, expanded along the drei-bein of the torus, $v_i^{\hat{\alpha}}$, and the $B$-field components expanded along the harmonic 1-forms: $b_i^\alpha$. For generic points in the moduli space the gauge group is the Cartan subgroup of $G$, $U(1)^{16}$, together with the six abelian gauge fields from metric and $B$-field. 

On the other hand, starting with M-theory on a K3-surface, we have the metric $\tilde{g}_{MN}$ and three-form $C_{MNK}$. The metric deformations account for the moduli space $\mathcal{M}$ by a choice of three complex structures $\omega_1, \omega_2, \omega_3$ in $\Gamma^{3,19} \otimes \mathbb{R}$, which precisely corresponds to the choice of a point in the Grassmanian $SO(3,19)/SO(3)\times SO(19)$. The lattice automorphisms, which preserve this oriented three-plane correspond to automorphisms of the K3-surface. 

The gauge fields are obtained as follows:
The K3 has 22 two-forms, along which the 3-form gives rise to abelian gauge fields ${\bf a}_{i}^{\kappa}$
\be
C_{3} = c_{ijk} dy^i \wedge dy^j \wedge dy^k +    \sum_{\kappa=1}^{22} {\bf a}_{i}^{\kappa} \omega^{(2)}_{\kappa} \,.
\ee
These $22$ $U(1)$ gauge fields match with the heterotic spectrum at generic points in moduli space. 
The three-form $c$ in 7d can be dualized to a two-form
\be
dc = \star_7 db\,,
\ee
which matches with the 2-from in the heterotic compactification. 

For specific points in the moduli space, there are extra massless states which correspond to enhanced gauge symmetry. From the heterotic side, this happens whenever there is a lattice point $\gamma \in \Gamma^{3,19}$ such that $\gamma^2 = -2$, which is perpendicular to the three-plane spanned by the $\omega_i$ in $\Gamma^{3,19}\otimes \R$. The set of all such $\gamma$ span a root lattice of ADE type, which corresponds directly to the unbroken non-abelian gauge group of the theory. On the M-Theory side, these points in moduli space are precisely those ones for which the K3 surface develops the corresponding ADE singularities. The gauge degrees of freedom arise from massless M2-branes wrapped on the collapsed rational curves. 
In a fiber-wise application of heterotic/M-theory duality, it hence becomes natural to find a geometry for which every K3-fiber is singular. It is precisely this behavior, which we will uncover for M-Theory on compact TCS $G_2$-manifolds  with heterotic duals.


\subsection{Fiber-wise Duality for Elliptic K3s}
\label{sec:FibwiseDuality}

The M-theory/heterotic duality will now be applied fiberwise over a three-dimensional base, which on the M-theory side results in a $G_2$-manifold and on the heterotic in a Calabi-Yau threefold  compactification. 
The TCS-construction for $G_2$-manifolds  of section \ref{sec:TCSG2} has a natural such K3-fibration, see (\ref{K3J}), which suggests that they are particularly suitable for the application of such a fiber-wise duality. We will show, that indeed the duality can be explicitly realized, when the K3-fibers of each TCS building block are elliptically fibered, i.e. 
\be
\mathbb{E} \ \hookrightarrow \ K3 \ \rightarrow \mathbb{P}^1 \,,
\ee
where the elliptic curve $\mathbb{E}$ is holomorphically fibered over the $\mathbb{P}^1$.
To implement this, it will be useful to keep the following geometric point of view in mind: 
We should think about the hyper-K\"ahler structure, i.e. the three harmonic self-dual two-forms $\omega_i$, of K3 as being related to the three $\Sone_i$ of $T^3$. In particular, we may expand $\omega_i$ in 
\be
H^2(K3,\Z) = U_1 \oplus U_2 \oplus U_3 \oplus (-E_8)^{\oplus 2}
\ee
and we can think of the periods of $\omega_i$ in $U_i$ as the radius of each $\Sone$ ($\omega_i$ in $U_j$ for $i\neq j$ are off-diagonal elements of the metric on $T^3$) and the periods of $\omega_i$ in $(-E_8)^{\oplus 2}$ as the Wilson lines around $\Sone_i$.

Note that the case where $\omega_3$ has non-trivial periods on one of the $U_E \subset H^2(K3,\Z)$ (together with some sublattice of $E_8^{\oplus 2}$) but $\omega_1,\omega_2$ are perpendicular to $U_E$ is particularly simple as this gives a factorization of $T^3 = T^2_f \times \Sone_3$. If we think of $\omega_3$ as the K\"ahler form, this condition is nothing but demanding that the K3-surface $S$ has an elliptic fibration with a section. If we fiber such an elliptic K3-surface $S$ holomorphically over $\P^1$ to get a Calabi-Yau three-fold, only $\Omega^{2,0} = \omega_1 + i \omega_2$ varies and $\omega_3$ stays constant, so that in the heterotic dual, we only fiber a $T^2_f \subset T^3$ non-trivially over $\P^1$ and we end up with heterotic on {an elliptic} K3$ \times \Sone_3$. Besides the twisting of $T^2_f = \Sone_1 \times \Sone_2$ over the base $\P^1$, the K3-fibration also determines a flat vector bundle on $T^2_f$ via a spectral cover. This is encoded in the (varying) periods of $\Omega^{2,0}$ within the $E_8^{\oplus 2}$ lattice over the $\P^1$ base of the building block. Furthermore, the integrals of the K\"ahler form $\omega_3$ over cycles associated with the $E_8^{\oplus 2}$ lattice correspond to a Wilson line on  $\Sone_3$. 

The picture of the duality between M-Theory and heterotic string theory we have painted here lifts to the duality between F-theory and heterotic string theory in six dimensions and leads to (a subset of examples for) the duality between type IIA string theory and heterotic string theory upon circle compactification.

Although the global picture for a $G_2$-manifold realized as a twisted connected sum is more complicated, we can use the above logic for each K3-fibered building block separately, apply the duality, and then glue the  `building blocks'  of the heterotic dual together as dictated by the matching \eqref{eq:DM}. This will imply that both the geometry and the bundle data are consistently identified on the heterotic side. Using only elliptic K3-surfaces as fibers on the building blocks
of the $G_2$-manifold \footnote{Due to the matching condition \eqref{eq:DM}, the elliptic fiber on one side is identified with a non-algebraic torus on the other side. } allows us to explicitely describe the dual heterotic geometry as it simplifies discriminating between geometry and bundle data. However, it is possible to apply the fiberwise duality for TCS building blocks more generally and consider non-elliptic K3-surfaces as fibers of building blocks. We leave an investigation of such models for the future. 

%
%

\section{Heterotic Dual for smooth TCS $G_2$-manifolds}
\label{sec:TCSSYZ}

We now apply the fiber-wise M-theory/heterotic duality to the TCS construction of $G_2$-manifolds for which the K3-fibers are elliptic, and determine a TCS-like construction of the dual heterotic compactification. 
The latter will be constructed from two building blocks, which are open hyper-K\"ahler manifolds, that are glued using the dual of the map in the $G_2$ TCS construction. 
Under M-theory/heterotic duality, the K3-fibration of the TCS $G_2$-manifold maps to the SYZ $T^3$-fibration of the dual heterotic Calabi-Yau threefold.
We show in particular, that the resulting Calabi-Yau three-fold is the Schoen manifold, i.e. the split bi-cubic, for all such TCS $G_2$-manifolds.

\subsection{TCS for SYZ Calabi-Yau Three-folds}

The starting point is M-Theory on a TCS $G_2$-manifold $J$. In the twisted connected sum construction for this $G_2$-manifold, we employ algebraic three-folds $Z_\pm$ as building blocks, as explained in section \ref{sec:TCSG2}. These are K3 fibrations with fiber $[S]$ over $\widehat{\P^1}$, such that the canonical class of $Z_\pm$ is given by a fiber $[S]$. The building blocks themselves are hence not Calabi-Yau (\ref{c1Z}).

The starting point for the heterotic three-fold $X_{\rm het}$ is as in section \ref{sec:FibwiseDuality}, where the duality is applied to the setup with an elliptically fibered K3. 
Recall that for an elliptic K3, that is in addition fibered over $\widehat{\mathbb{P}}^1$, only two of the three complex structures, $\omega_1$ and $\omega_2$, of the K3 vary. In the heterotic dual model, by fiberwise duality, therefore only a $T^2_f \subset T^3$ varies over the base $\widehat{\mathbb{P}}^1$, and the total space of the heterotic compactification is an elliptic K3$\times \Sone_3$. We now apply the same ideas in the TCS construction to this Calabi-Yau three-fold.

First we twist the fiber so as to have a non-trivial first Chern-class as in (\ref{c1Z}), i.e. the resulting geometry is $dP_9 \times \Sone_3$ with 
\be
c_1 (dP_9) = [T^2_f]  \,,
\ee
where $[T^2_f]$ is the class of the torus fiber $T^2_f = \Sone_1 \times \Sone_2$. 
The next step in the TCS for $G_2$-manifolds is the removing a central fiber from $Z_\pm$ to form $X_\pm = Z_\pm \setminus S_{0\pm}$. Applying the same process to the heterotic model, we define two building blocks\footnote{We use the term building block here in  analogy to the TCS construction, as $V_\pm$ play a similar role as $Z_\pm$, applied to the construction of the Calabi-Yau three-fold.} 
\be
V_\pm = dP_9 \setminus T^2_f\,.
\ee

\begin{figure}
\begin{center}
  \includegraphics[height=6cm]{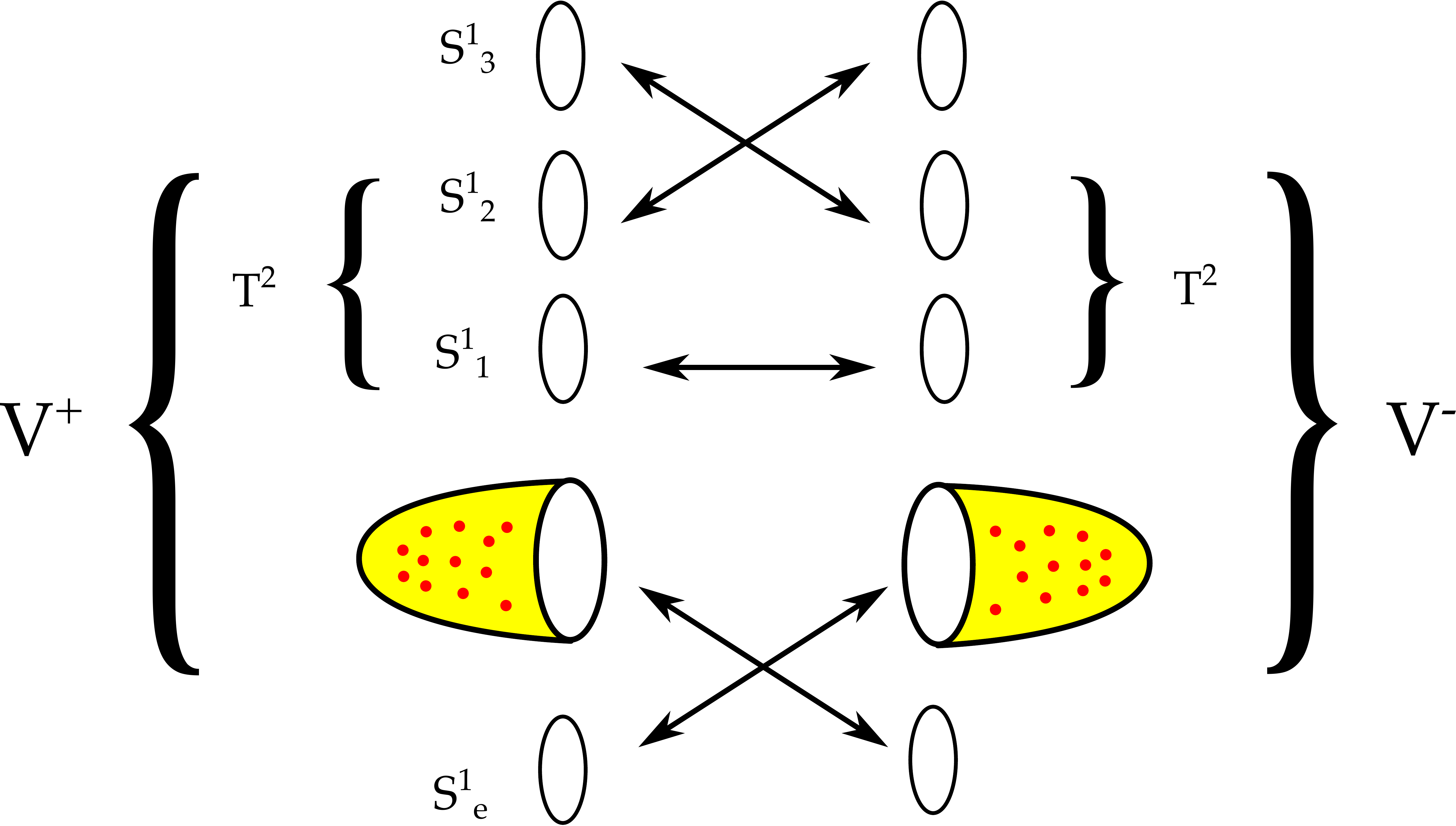}
 \caption{\label{fig:glue} A cartoon of the Calabi-Yau three-fold $X_{\rm het}= X_{19,19}$ glued from two copies of {$V_\pm\times T^2_\pm$ where $T^2_\pm = \Sone_3 \times \Sone_e$}. The elliptic fiber of $V$ degenerates over $12$ points of the open $\P^1$ base. }
\end{center}
\end{figure}

From the duality, the $V_\pm$ must asymptote to $I \times \Sone_b \times \Sone_1 \times \Sone_2$.
In the $G_2$ construction, the Donaldson matching identifies
\begin{equation}
\begin{aligned}
\omega_{1+} &= \omega_{3-} \\
\omega_{2+} &= - \omega_{2-} \\
\omega_{3+} &= \omega_{1-}\,,
\end{aligned}
\end{equation}
which in the dual heterotic Calabi-Yau now becomes the identification
\begin{equation}\label{eq:T3gluing}
\begin{aligned}
\Sone_{1+} &= \Sone_{3-} \\
\Sone_{2+} &= \Sone_{2-} \\
\Sone_{3+} &= \Sone_{1-} \,.
\end{aligned}
\end{equation}
Finally, the twisted connected sum uses $X_\pm \times \Sone_{e\pm}$ and glues the $\Sone_{b\pm}$ in the open $\P^1 \setminus p_0$ bases of $X_\pm$ to {$\Sone_{e\mp}$}, 
\be\label{eq:Sonebe}
{\Sone_{e\pm}  = \Sone_{b\mp}\,,}
\ee
which we may equally well apply to $\Sone_3 \times V_{\pm}$. In conclusion, the Calabi-Yau three-fold is obtained by gluing 
\be
M_\pm = V_{\pm}\times T^2_\pm \,,\qquad T^2_\pm =  \Sone_{3\pm}\times\Sone_{e\pm} \,,
\ee
along 
\be
W = M_+ \cap M_- \cong I \times (\Sone)^5 \,,
\ee
with the  identifications in (\ref{eq:T3gluing}) and (\ref{eq:Sonebe}) -- see figure \ref{fig:glue} for a cartoon.
We denote the resulting Calabi-Yau three-fold by $X_{\rm het}$.

To identify the specific Calabi-Yau, first note that the Euler characteristic of $X_{\rm het}$ is given by
\begin{equation}
 \chi(X_{\rm het}) = \chi(M_+) + \chi(M_-) - \chi(W) =0 \, ,
\end{equation}
which follows immediately, from additivity of the Euler characteristic. We used furthermore that $M_+$, $M_-$ and $W$ all have an $\Sone$-factor, and thereby vanishing  Euler characteristic. Next, we may use the Mayer-Vietoris sequence to compute the Betti numbers of $X_{\rm het}$ as 
\begin{equation}
H^m(X_{\rm het},\Z) = \ker(\gamma^m) \oplus \coker(\gamma^{m-1}) \,,
\end{equation}
where
\begin{equation}
\gamma^m : \qquad H^m(M_+\Z) \oplus H^m(M_-,\Z) \rightarrow H^m(W,\Z)\,.
\end{equation}
It is not hard to see that $\ker(\gamma^1)=0$, which implies that $H^1$ is trivial. 
Futhermore, $|\coker(\gamma^2)|=3$ and $|\ker(\gamma^2)| = 16$, so that $h^{1,1}(X_{\rm het}) =19$, and thereby $h^{2,1} (X_{\rm het})=19$ as well, i.e. $X_{\rm het} = X_{19,19}$. This Calabi-Yau is known as the Schoen manifold \cite{Schoen} or the split bi-cubic.

Note that \emph{any} M-theory compactification on a TCS $G_2$-manifold of the type considered here, i.e. with elliptic K3-fibers, is dual to a heterotic compactifications on $X_{19,19}$. The multitude of such TCS $G_2$-manifolds simply corresponds to different choices of vector bundles on $X_{19,19}$. 

\begin{figure}
\begin{center}
  \includegraphics[height=3cm]{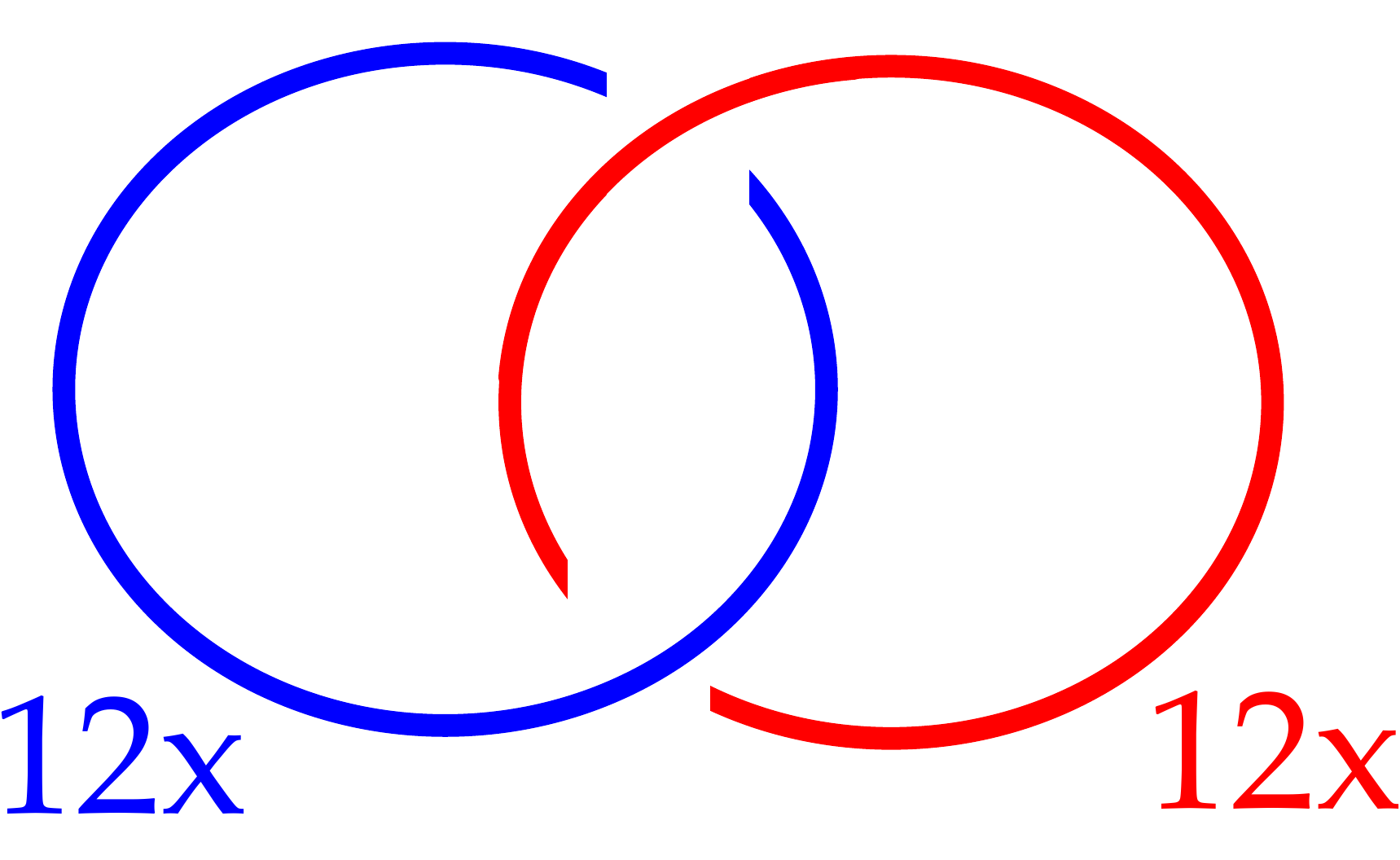}
 \caption{\label{fig:hopf} The Hopf link. The discriminant locus of the TCS fibration is a collection of 12 such Hopf links in the base of the fibration, which are retract to the above image.}
\end{center}
\end{figure}

The SYZ-fibration of $X_{\rm het}$ is known and has a simple structure \cite{Gross}, which we will recover from the M-Theory duals: the discriminant locus consists of two sets of twelve disjoint $\Sone$'s. Any two $\Sone$'s from different groups have linking number $1$ (Hopf link) and $0$ otherwise. The same can be recovered from the Kovalev construction given above: the discriminant locus of the SYZ-fiber is $\Sone$ times the points over which the elliptic fiber of $V$ degenerates. As the two building blocks $V_\pm$ are open versions of $dP_9$, this happens over 12 points each. The $\Sone$ in the discriminant locus is the same as $\Sone_{e\pm}$ on both sides. As these are swapped with $\Sone_b$, the two groups of $12$ $\Sone$'s forming the discriminant locus are interlocked like the Hopf links shown in figure \ref{fig:hopf}.

\subsection{Various Description of the Schoen Calabi-Yau}\label{sect:prettylittlething}

In this section we give various constructions and limits of the heterotic Calabi-Yau three-fold $X_{\rm het}= X_{19,19}$, i.e. the Schoen manifold, which will be useful in the following.

\subsubsection{Orbifold Limit}

The Calabi-Yau with Hodge numbers $X_{19,19}$ is a rather well-studied. It can be found as a resolution of the orbifold 
\be
X_o = T^6/(\Z_2\times\Z_2)\,,
\ee
 where the $\Z_2^2$ acts as
\begin{equation}
\begin{aligned}
\alpha: (z_1,z_2,z_3)& \rightarrow (z_1,-z_2,-z_3) \\
\beta: (z_1,z_2,z_3) &\rightarrow (-z_1,z_2,-z_3+\tfrac12) \\
\alpha\beta: (z_1,z_2,z_3) &\rightarrow (-z_1,-z_2,z_3-\tfrac12) \,.
\end{aligned}
\end{equation}
Let us try to see this explicitely. First note that we can think of $V = dP_9\setminus T^2_f$ as an elliptic fibration over an open $\P^1$ with two $I_0^*$ fibers, i.e.
\begin{equation}
( T^2_f \times \Sone_b \times \R ) / \Z_2 \,,
\end{equation}
where the $Z_2$ acts by giving a minus sign to all four directions. Note that the above gives a K3-surface when we replace $\R$ by $\Sone$, which allows us to see that a K3-surface  could be glued from two copies of the building blocks $V$. 

We can now describe the above orbifold in detail. Let us choose coordinates
\begin{equation}
\begin{aligned}
 z_1&=x_1 + i x_2\\
 z_2&=x_3+ix_4\\
 z_3&=x_5+ix_6\,,
\end{aligned}
\end{equation}
so that the orbifold action is
\begin{equation}
\begin{aligned}
  \alpha:& (x_1,x_2,x_3,x_4,x_5,x_6) \mapsto (x_1,x_2,-x_3,-x_4,-x_5,-x_6)\\
 \beta:  & (x_1,x_2,x_3,x_4,x_5,x_6)  \mapsto (-x_1,-x_2,x_3,x_4,-x_5,-x_6+\tfrac12) \, .
\end{aligned}
\end{equation}
We can pull apart this orbifold along the $x_6$ direction. Note that any $x_6$ can be mapped to the closed interval $[0:\tfrac14]$ by $\alpha$ and $\beta$, with $x_6 = 0$ being fixed by $\alpha$ and $x_6=\tfrac14$ fixed by $\beta$. Hence we can think of $X_o$ as being fibered over $x_6=[0:\tfrac14]$ with generic fiber $(\Sone)^5$ and `ends' $M_+ = (\Sone)^5\times \R/\Z_2^\alpha$ and $M_- = (\Sone)^5\times \R/\Z_2^\beta$. 

The coordinates $x_i$ are associated with the various ingredients in figure \ref{fig:glue} as follows. In the vicinity of $x_6=0$, the involution $\alpha$ must act with a sign on $T^2_f = \Sone_1\times \Sone_2$ and $\Sone_b$ of $M_+$, together with the interval direction $x_6$. Similarly, in the vicinity of $x_6=\tfrac14$, the involution $\beta$ must act with a sign on $T^2_f = \Sone_1\times \Sone_2$ and $\Sone_b$ of $M_-$. This results in the identification\footnote{Note that this choice is not unique, in fact $x_1 \leftrightarrow x_2$ and $x_3 \leftrightarrow x_4$ are symmetries of the whole configuration. We have made a choice for which our SYZ fiber is sLag.}
\begin{equation}
M_+\,\, \left\{ \begin{array}{ccc}
                  \Sone_1 \sim &x_1& \sim \Sone_3 \\
                  \Sone_2 \sim &x_5& \sim \Sone_2 \\
                  \Sone_3 \sim &x_3&\sim \Sone_1 \\
                  \Sone_b \sim &x_2& \sim \Sone_e \\
                  \Sone_e \sim &x_4& \sim \Sone_b 
                \end{array}\right\} \,\, M_-
\end{equation}
and reproduces the gluing \eqref{eq:T3gluing}, so that $X_o$ is decomposed precisely in the same way as shown in figure \ref{fig:glue}. This decomposition has been considered by A.Kovalev (unpublished); although he conjectured the existence of a Ricci-flat Calabi-Yau metric on this twisted connected sum Calabi-Yau, this has not yet been proven rigorously. 

The above identification allow us to see the holomorphic coordinates of $X_{\rm het}$ as realized by gluing $M_\pm$. In the identification above, the $T^3$ in $X_{\rm het}$ which replaces the K3 on the $G_2$-manifold $J$ is given by $\Im(z_1) = \Im(z_2) = \Im(z_3) = 0$, so it is special Lagrangian.

Let us now describe how the classic presentation of the Schoen Calabi-Yau $X_a$ as the split bi-cubic is related to the orbifold $X_o$. The Schoen Calabi-Yau is realized as a CICY in $\P^2\times\P^2\times \P^1$ with configuration matrix, which indicates the degrees of the hypersurfaces
\begin{equation}\label{eq:configschoen}
 \left[\begin{array}{c|cc}
\mathbb{P}^2 & 3 & 0 \\
\mathbb{P}^2&  0 & 3 \\
\mathbb{P}^1&  1 & 1
 \end{array}\right] \,.
\end{equation}
This can be visualized in various ways. First note that, fixing a point on the $\P^1$, the complete intersection \eqref{eq:configschoen} becomes a product of two elliptic curves. We may consider projections which forget one of the $\P^2$ factors of the ambient space. These project to hypersurfaces of degree $(3,1)$ in $\P^2 \times \P^1$, which are nothing but a rational elliptic surfaces $dP_9$. The inverse image of such a projection is an elliptic curve. Hence  $X_a$ can be thought of as an elliptic fibration over $dP_9$ in two different ways. Finally, note that both of these $dP_9$ surfaces share the same base as elliptic surfaces, which implies that 
\begin{equation}
 X_a = dP_9 \times_{\P^1} dP_9 \, .
\end{equation}
The manifold \eqref{eq:configschoen} has second Chern class
\begin{equation}\label{eq:c2schoen}
 c_2(X_a) = 3\left(H_1^2 + H_2^2 + H_1 \cdot H_3 + H_2\cdot H_3 \right) \,,
\end{equation}
where $H_1,H_2$ are the hyperplane classes of the $\P^2$ factors and $H_3$ is the hyperplane class of the $\P^1$ factor of the ambient space. As $H_1^2$ fixes a unique point on the first $\P^2$, which then gives a unique point on the $\P^1$ via the first equation, $H_1^2$ corresponds to a single elliptic curve. Similarly, $ H_1 \cdot H_3$ gives three copies of the same elliptic curve. We hence find that we can represent $c_2(X_a)$ as $12$ copies of each of the elliptic curves corresponding to the fibers of the two elliptic fibrations. 

The same structure can be seen from the orbifold version $X_o$. Here, we identify $z_3$ with the base and the projection to each of the $dP_9$ is given by forgetting the directions $z_1$ or $z_2$. The image of the projection is $(T^4/\Z_2^k)/\Z_2^s$. Here, $\Z_2^k$ acts by inverting all coordinates producing a Kummer surface, and $\Z_2^s$ pairwise identifies the four $I_0^*$ fibers of the Kummer surface see as an elliptic fibration. This produces a $dP_9$ as an elliptic surface with two $I_0^*$ fibers.  

\begin{figure}
 \begin{center}
  \includegraphics[height=6cm]{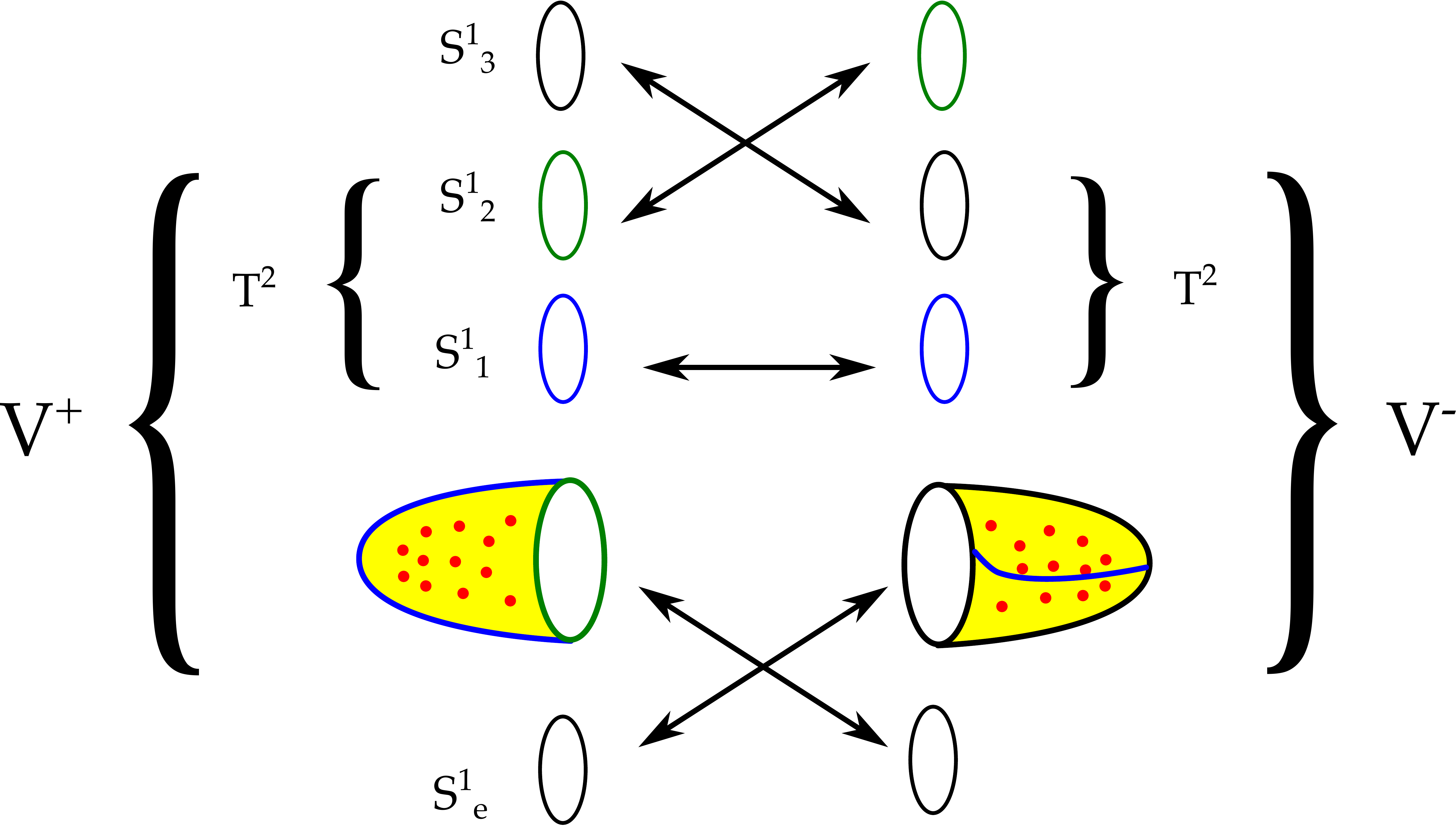}
  \caption{\label{cydecomp2} The Kovalev limit of the Schoen Calabi-Yau three-fold. We have shown one of the two $dP_9$s in blue and green: it is the one arising from a projection to $z_1,z_3$. As the $dP_9$ includes all of $V_+ = dP_9\setminus T^2_f$, one can see the $12$ singular fibers. On $M_-$, the circle factors experience no more monodromies. Furthermore, the elliptic fiber of the $dP_9$, which becomes one of the elliptic fibers of $X_{\rm het}$ is colored in green. It corresponds to the coordinate $z_1$ of the orbifold. The other $dP_9$ originating from projection to $z_2,z_3$ is found by swapping $M_+ \leftrightarrow M_-$.}
 \end{center}
\end{figure}

The above allows to identify the two $dP_9$'s and the elliptic fiber in our Kovalev picture, we have shown one such choice in figure \ref{cydecomp2}.

Although we should think of the $T^3 = \Sone_1 \times \Sone_2 \times \Sone_3$ as an SYZ fiber $T^3_{SYZ}$ of the Calabi-Yau $X_{\rm het}$, the $T^2_f \subset T^3_{SYZ}$ is seemingly holomorphic as the fiber of $dP_9$. However, the holomorphic directions are different, it is $z_1$ of $M_+$ which is the holomorphic coordinate on the fiber $\Sone_1 \times \Sone_b$ of one $dP_9$ ($z_2$ of $M_+$ is the coordinate on the elliptic fiber $\Sone_1 \times \Sone_b$ of the other) and $z_3$ is the coordinate on $\Sone_2 \times x_6$, which is the base. Recall that $x_6$ is the direction of the interval along which we do the Kovalev decomposition of $X_a$.

\subsubsection{Realization as an Elliptic Fibration with Section}\label{sect:schoentoric}

The Schoen Calabi-Yau $X_{\rm het}$ can also be realized by a Weierstrass model over $dP_9$, which can itself be realized as an elliptic surface over $\P^1$. This can be explicitely constructed as a complete intersection in an ambient toric variety with weight system 
\begin{equation}\label{eq:weightsschoentoric}
 \begin{array}{rrrrrrrr|c|rr}
\hat{z}_1 & \hat{x} & \hat{y} & x & y & w & \hat{w} & \hat{z}_2 & \Sigma \hbox{ of degrees} & W&\hat{W} \cr \hline 
1 & 2 & 3 & 2 & 3 & 0 & 0 & 1 & 12 & 6 & 6 \\
0 & 2 & 3 & 0 & 0 & 0 & 1 & 0 & 6  & 0 & 6 \\
0 & 0 & 0 & 2 & 3 & 1 & 0 & 0 & 6 & 6 & 0
 \end{array}
\end{equation}
with the defining equations
\begin{equation}\label{eq:eqschoentoric}
\begin{aligned}
W &= -y^2 + x^3 + f_4(\hat{z}_1,\hat{z}_2) x w^4 + g_{6}(\hat{z}_1,\hat{z}_2) w^6 & = 0 \\
\hat{W} &= -\hat{y}^2 + \hat{x}^3 + \hat{f}_4(\hat{z}_1,\hat{z}_2) \hat{x} \hat{w}^4 + \hat{g}_{6}(\hat{z}_1,\hat{z}_2) \hat{w}^6 & = 0 \,,
\end{aligned}
\end{equation}
where $f,g,\hat{f},\hat{g}$ are homogeneous polynomials of the indicated degrees in $[\hat{z}_1:\hat{z}_2]$. Using \cite{Batyrev:1994pg} (see Appendix \ref{app:nef} and \ref{app:schoentoric}), the Hodge numbers turn out to be $h^{1,1}(X_{\rm het}) = h^{2,1}(X_{\rm het}) =19$. 

Projecting onto the coordinates $\hat{y},\hat{x},\hat{w},\hat{z}_1,\hat{z}_2$ realizes one elliptic fibration of $X_{\rm het}$. The base $B_s$ of this elliptic fibration is a $dP_9$ given by $\hat{W} = 0$. The elliptic fibration of $B_s$ is in turn found by projecting to $x, y, w, \hat{z}_1,\hat{z}_2$, i.e. we forget the coordinates $\hat{x},\hat{y},\hat{w}$. Note that the whole three-fold $X_{\rm het}$ can also be seen as being fibered by two elliptic curves $\mathbb{E}$ and $\widehat{\mathbb{E}}$ over the $\widehat{\P}^1$ with coordinates $[\hat{z}_1:\hat{z}_2]$ and that there is a second elliptic fibration found by swapping the roles of $\hat{y},\hat{x},\hat{w}$ and $x,y,w$. A cartoon can be found on the right hand side of figure \ref{fig:XF}.

We can understand $h^{1,1}(X_{\rm het})=19$ as follows: there are $10$ algebraic cycles in $h^{1,1}(dP_9)$ for each of the $dP_9$s, but the two $dP_9$s share a common base (the $\widehat{\P}^1$ with coordinates $[\hat{z}_1:\hat{z}_2]$). If we choose to consider the elliptic fibration associated with $x,y,w$, the base $dP_9$ $\hat{W} = 0$ in $\hat{y},\hat{x},\hat{w},\hat{z}_1,\hat{z}_2$ has $h^{1,1}(B_s)=10$. Crucially, the equation $W=0$ does not depend at all on the coordinates $\hat{x},\hat{y},\hat{w}$, i.e. the elliptic fibration is trivial over the elliptic curve $\widehat{\mathbb{E}}$. 

In this presentation, the second Chern class of $X_{\rm het}$ is given by
\begin{equation}\label{eq:c2weierXa}
 c_2(X_{\rm het}) = 12\, [\hat{z}_1] \cdot ([w]+[\hat{w}]) = 12 (\widehat{\mathbb{E}} + \mathbb{E})\, .
\end{equation}
Here $\mathbb{E}$ and $\widehat{\mathbb{E}}$ are the curve classes represented by the two elliptic curves. 
%
%

\subsection{Dual Smooth TCS $G_2$-manifolds}\label{eq:g2hetvanilla}

In this section we study a simple compactification of heterotic $E_8 \times E_8$ string theory on $X_{\rm het}$ and its $G_2$ dual. For this, we consider a generic $E_8\times E_8$ bundle $V$ with $ch_2(V) = 12 \widehat{\mathbb{E}}$ in which the instantons are distributed as $(6,6)$ between the two $E_8$ factors. In this case, each $E_8$ bundle has $112$ moduli and hence contributes $112$ chiral multiplets to the 4d effective field theory \cite{Curio:1997rn}. As we furthermore need to satisfy the anomaly condition
\begin{equation}
 ch_2(X_{\rm het}) = ch_2(V) + [\hbox{NS5}]
\end{equation}
we need to introduce $12$ NS5-branes wrapped on the elliptic curve $\mathbb{E}$.
Each of these NS5 branes gives a $U(1)$ vector multiplet and three chiral multiplets in four dimensions. Together with the geometric moduli and the dilaton, we hence find 
\begin{equation}\label{hetspec}
\begin{aligned}
 n_v &= 12  \\
 n_c &= 1+ 2\cdot 19 + 2\cdot 112  + 36 = 299
 \end{aligned}
\end{equation}
for the number of vector multiplets ($n_v$) and chiral multiplets ($n_c$). 

We now want to reproduce the above from a TCS $G_2$-manifold. As we want a generic bundle in $E_8$ on one of the two elliptic curves $\mathbb{E}$ and $\widehat{\mathbb{E}}$, it seems natural to construct a pair of building blocks for which the generic K3-fiber has the lattices
\begin{equation}
\begin{aligned}
T_+ &=&  E_8 \oplus E_8 \oplus U^{\oplus 2} &\hspace{1cm} &&N_+ = U \\
T_- &=&  U^{\oplus 2} &&&N _- = E_8 \oplus E_8 \oplus U\, .
\end{aligned}
\end{equation}
We naturally have an orthogonal gluing with $N_+ \cap N_- = 0$ under the obvious identification. Furthermore, we expect to be able to choose the hyper-K\"ahler structure  on both sides such that the matching \eqref{eq:DM} can be satisfied.  While both K3 fibrations together give rise to the non-trivial geometry of $X_{\rm het}$, only $Z_+$ carries the information of the $E_8\times E_8$ bundle!

Let us be more concrete. For the building block $Z_+$, let us take a hypersurface in a toric variety with weight system
\begin{equation}
\begin{array}{ccccccc|c}
y & x & w & z_1 & z_2 & \hat{z}_1 & \hat{z}_2 & P \\
\hline
3 & 2 & 1 & 0 & 0 & 0 & 0 & 6\\
6 & 4 & 0 & 1 & 1 & 0 & 0 & 12\\
3 & 2 & 0 & 0 & 0 & 1 & 1 & 6
\end{array}\, ,
\end{equation}
defined by an algebraic equation
\begin{equation}\label{eq:inst66model}
 P = -y^2 + x^3 + f_{8,4}(z,\hat{z}) x w^4 + g_{12,6}(z,\hat{z}) w^6 = 0\, .
\end{equation}
Here $y,x,w$ form a weighted projective space $\P^2_{123}$ and $f,g$ are polynomials of the indicated degrees in the coordinates $[z_1:z_2]$ and $[\hat{z}_1 :\hat{z}_2]$ on $\P^1 \times \widehat{\P^1}$. The building block $Z_+$ is then a generic fibration of a Weierstrass elliptic K3-surface over $\widehat{\P}^1$. 
In the language of tops, this is recovered from 
\begin{equation}
 \Diamond^\circ_+ = \left(\begin{array}{rrrrr}
-1 & 0 & 2 & 2 & 2 \\
0 & -1 & 3 & 3 & 3 \\
0 & 0 & -1 & 1 & 1 \\
0 & 0 & 0 & 0 & 1
\end{array}\right) 
\end{equation}
using the construction reviewed in Section \ref{sec:TCSG2}.
With \eqref{eq:bbtopology}, we compute
\begin{equation}
h^{1,1}(Z_+) =  3 \hspace{1cm} h^{2,1}(Z_+) =  112 \hspace{1cm} |N(Z_+)| = 2 \, ,
\end{equation}
for this space, so that $K(Z_+) = 0$ follows. 

For $Z_-$, we use K3-surfaces in the family with $N = U \oplus (-E_8)^{\oplus 2}$ as the fibers. This can be done by a choice
\begin{equation}\label{eq:bbe8e8}
\Diamond^\circ_- = \left(\begin{array}{rrrrr}
-1 & 0 & 2 & 2 &  2 \\
0 & -1 & 3 & 3 &  3 \\
0 & 0 & -6 & 0 &  6 \\
0 & 0 & 0 & 1 &  0
\end{array}\right)
\end{equation}
which realizes a generic fibration of a resolved Weierstrass K3 with two fibers of type $II^*$ over a $\P^1$. This space has
\begin{equation}\label{eq:databb-}
h^{1,1}(Z_-) =  31 \,, \hspace{1cm} h^{2,1}(Z_-) =  20\,, \hspace{1cm} |N(Z_-)| = 18 \,,\hspace{1cm} |K(Z_-)| = 12 \,.
\end{equation}

For an orthogonal gluing of the above building blocks, we hence find a smooth $G_2$-manifold $J$ with Betti numbers
\begin{equation}\label{eq:bettig2DGW}
\begin{aligned}
 b_2(J) &= 12 \\
 b_3(J) &= 23 + 2(112+20) + 12 = 299 \,,
\end{aligned}
\end{equation}
which precisely reproduces the spectrum found for the heterotic compactification in (\ref{hetspec}),  by identifying $b_2(M)$ with $n_v$ and $b_3(M)$ with $n_c$! 

This identification of spectra is a very strong indication that for the smooth TCS $G_2$-manifolds with elliptic K3 building blocks we have identified the heterotic dual compactifications. In the remainder of this paper we will generalize this construction of dual models, culminating in a general proof of the equivalence of spectra in section \ref{sec:GeneralStory}. 
This includes models with non-abelian gauge symmetries, which in M-theory correspond to singular TCS $G_2$-manifolds.


\section{M-theory on $G_2$ with Non-Higgsable Gauge Groups}
\label{sec:NHC}

In this section, we will generalize the construction of dual models of the last section by considering models with different distributions of instanton numbers. In particular, this will give models with non-Higgsable gauge groups, i.e. $G_2$-manifolds with non-deformable singularities. The generalization to include Higgsable gauge groups is provided in section \ref{sec:GeneralStory}.

\subsection{Distinguishing between Geometry and Bundles}
\label{sec:GeoBun}

To prepare our discussion, let us examine the $G_2$-manifold discussed in the last section more closely. Over the base of each building block $Z_\pm$, the K3-fiber undergoes monodromies and degenerates over a number of points. These monodromies correspond to the monodromies of the SYZ fibration of the Schoen Calabi-Yau, together with the $E_8 \times E_8$ bundle on this geometry.  A degeneration of the K3-fiber of a $G_2$-manifold into a (small) $T^3$ which is almost constant over an interval (with non-trivial behavior only happening at the ends) corresponds to the limit in which the SYZ $T^3$ on the heterotic side becomes large and we have a semiclassical description. It is precisely this limit which allows us to distinguish between geometry and bundles from the perspective of the K3-surface: the monodromies of the $T^3$ in the bulk of the interval give us the monodromies of the SYZ fiber and the monodromies affecting the cycles in the two ends correspond to the bundle data \cite{DaveStrings2002}. As we have constructed our building blocks to be fibered by elliptic K3-surfaces, we can consistently split the $T^3$ over the whole building block into a $T^2_f$ (the elliptic curve) times an $\Sone$, which sits in the base of the elliptic K3. The degeneration into a $T^3$ over an interval can then be done in two separate steps, with only the $T^2$-part being non-trivial. This is, however, the same limit relevant for the duality between F-Theory and heterotic string theory, i.e. the well-known stable degeneration limit \cite{Friedman:1997yq}. Here, the K3-surface degenerates into two $dP_9$ surfaces which meet along a common elliptic curve, which is $T^2_f$. It is this curve which tracks the non-trivial behavior of $T^2 \subset T^3$ (the SYZ fiber), whereas the two $dP_9$ surfaces determine the two $E_8$ bundles. 

Let us make the above explicit for the two building blocks constructed in the last section. For $Z_+$, which is given by \eqref{eq:inst66model}, the K3-fiber aquires $A_1$ singularities over a number $n_\mu$ of points in the base $\widehat{\P}^1$ with coordinates $[\hat{z}_1,\hat{z}_2]$, without causing singularities in the three-fold $Z_+$. Whereas a generic fiber contributes $24$ to the Euler characteristic, such singular fibers contribute only $23$, so that we can compute $n_\mu$ for any smooth K3-fibered manifold $Z$ without reducible fibers from 
\begin{equation}
\chi(Z) = 24 (2-n_\mu) + 23 n_\mu = 48 - n_\mu  \, .
\end{equation}
As $\chi(Z_+) = -216$ we find $n_\mu(Z_+) = 264$. Applying the stable degeneration limit $S_+\rightarrow dP_9 \amalg dP_9$ for every fiber of $Z_+$, gives us a degeneration of $Z_+ \rightarrow \check{Z}_+ \amalg \check{Z}_+$. The elliptic curve in which the two $dP_9$ surfaces meet has a discriminant
\begin{equation}
\Delta= 4 f^3(\hat{z}) +  27 g^2(\hat{z}) \, ,
\end{equation}
which is a homogeneous polynomial of degree $12$. Hence there are $12$ monodromy loci for the SYZ fiber coming from $Z_+$. Similarly, the degeneration $S_-\rightarrow dP_9 \amalg dP_9$ for every fiber of $Z_-$ produces a degeneration $Z_- \rightarrow \check{Z}_- \amalg \check{Z}_-$. Again, the $T^2$ in which the two $dP_9$ fibers meet degenerates over $12$ points of the $\widehat{\P}^1[\hat{z}_1:\hat{z}_2]$. In the associated $G_2$-manifold $J$, each of these degeneration loci becomes an $\Sone$ and, due to the gluing between $Z_+ \setminus S_+$ and $Z_- \setminus S_-$, the $12$ circles from $Z_+$ are interlocked with the $12$ circles coming from $Z_-$. This precisely reproduces the structure of the SYZ fibration of the Schoen manifold reviewed in Section \ref{sect:prettylittlething}. In fact, it is not hard to see that this happens for any pair of building blocks fibered by elliptic K3-surfaces, which fits with the fact that they all correspond to compactifications of heterotic strings on $X_{\rm het}$ with different bundles.

To find the degeneration loci of the K3-fiber associated with the $E_8$ bundles, we have to find the number $\check{n}_\mu$ of degenerations of each of the two $dP_9$ fibers in $\check{Z}_\pm$ over $[\hat{z}_1,\hat{z}_2]$. This can again be done by computing their Euler characteristics and noting that a smooth $dP_9$ fiber contributes $12$ to the Euler characteristic while one with an $A_1$ singularity contributes $11$. We can hence write
\begin{equation}
 \chi(\check{Z}) = 12 (2-\check{n}_\mu) + 11 \check{n}_\mu = 24 - \check{n}_\mu\, .
\end{equation}
It is straightforward to compute $\chi(\check{Z}) = -96$, so that $\check{n}_\mu = 120 $ follows. 

Putting it all together, we have found that out of the $264$ degenerations of the K3-fiber of the building block $Z_+$, there are $120$ corresponding to each of the two $E_8$ bundles. The remaining $24$ monodromy points pairwise coincide in the limit in which the K3-surface degenerates into two $dP_9$ surfaces, these are the $12$ degeneration points of the SYZ fibration. 

In the other building block $Z_-$, each of the elliptic K3-surfaces it is fibered by has $E_8 \times E_8$ in its Picard lattice. There are no monodromies corresponding to bundle data, which is constant over the $\widehat{\P}^1$ base. There are only $24$ monodromy loci, which again pair up to form $12$ points in the degeneration limit of the K3-surface. These $12$ points correspond to the SYZ fibration on the heterotic side. 

\subsection{Instanton Configurations and $G_2$-manifolds}

With the detailed understanding of the monodromies of the K3 fibration gained in the last section, we have gained more confidence that the $G_2$-manifold $J$ constructed in Section \ref{eq:g2hetvanilla} indeed corresponds to a heterotic model on $X_{\rm het}$ with two $E_8$ vector bundles $V_I$ that has a distribution of instantons such that $ch_2(V_1) = ch_2(V_2) = 6 \widehat{\mathbb{E}}$. 

Note that the building block $Z_+$ we have employed is also elliptically fibered over $\mathbb{F}_0 = \P^1 \times \widehat{\P}^1$. From the similarity to the well-understood case of the duality between heterotic string theory and F-theory, it is natural to assume that trading $Z_+$ for a building block $Z_{+,n}$ which is elliptic over $\mathbb{F}_n$ corresponds to a distribution of instanton numbers
\begin{equation}\label{eq:instdistro}
\begin{aligned}
 \hbox{ch}_2(V_1) &= (6+n)\widehat{\mathbb{E}} \\
 \hbox{ch}_2(V_2) &= (6-n)\widehat{\mathbb{E}}
\end{aligned}
\end{equation}
This means we are interested in construction building blocks as hypersurfaces in a toric variety with weight system
\begin{equation}\label{eq:weightsysfnbb}
\begin{array}{ccccccc|c}
y & x & w & z_1 & z_2 & \hat{z}_1 & \hat{z}_2 & P \\
\hline
3 & 2 & 1 & 0 & 0 & 0 & 0 & 6\\
6 & 4 & 0 & 1 & 1 & 0 & 0 & 12\\
3+3n & 2+ 2n & 0 & 0 & n & 1 & 1 & 6+6n\\
\end{array}
\end{equation}
for $n = 0, \cdots, 6$. Let us first consider the case $n=1$, which gives a smooth hypersurface of Euler characteristic $\chi(Z_{+1}) = -216$ as well. The $240$ degeneration loci of the K3-fiber associated with the two $E_8$ bundles, however, are now distributed as $240 = 60 +180$. As before, this is found by degenerating all K3-fibers into $dP_9 \amalg dP_9$ and couting the number of degenerations of each of the $dP_9$ surfaces. 

For $n\geq 2$, a generic hypersurface $P_n=0$ in the ambient space defined by \eqref{eq:weightsysfnbb} above is singular and requires resolution. This can be seen as follows: the defining equation of $Z_{+,n}$ is just given by a Weierstrass model over $\mathbb{F}_n$
\begin{equation}
 P_n = -y^2 + x^3 + f_{8,4+4n}(z,\hat{z}) x w^4 + g_{12,6+n}(z,\hat{z}) w^6 = 0\, .
\end{equation}
where the subscripts indicate the weights of $f$ and $g$ under the scalings \eqref{eq:weightsysfnbb}. This means that $f$ and $g$ necessarily vanish over $z_1=0$ for $n\geq 2$ and the elliptic fibration has a corresponding singularity. The vanishing degrees of $f,g$ and the discriminant $\Delta$, as well as the ADE group and Kodaira fiber type associated with the degeneration, are given by
\begin{equation}\label{eq:nhcbb}
 \begin{array}{c|ccccc}
 n & f & g & \Delta & \mbox{fiber type} & G\\
 \hline
1 & 0 & 0 & 0& \mbox{smooth} & - \\
2 & 2 & 3 & 6& I_0^*& D_4 \\
3 & 3 & 4 & 8& IV^*& E_6\\
4 & 3 & 5 & 9& III^*& E_7\\
5 & 4 & 5 &10& II^* & E_8\\
6 & 4 & 5 &10& II^* & E_8
\end{array}\, .
\end{equation}
As these models correspond to heterotic compactifications with the instanton distributions \eqref{eq:instdistro}, we expect to find an unbroken gauge group whenever $6-n$ becomes too small to fully break $E_8$. Although the computation of the unbroken gauge group together with the spectrum can be done directly in the dual heterotic models, we find it more convenient to confirm the appearance of the above groups from the F-theory perspective. This is done in Section \ref{sect:F-theory}.

Using the description in terms of projecting tops, the resolutions of these spaces can easily be constructed and analysed, see Appendix \ref{app:bb} for the details. The resulting smooth building blocks $Z_{+,n}$ satisfy
\begin{equation}\label{eq:databbn}
 \begin{array}{c|ccc}
  n & |N(Z_{+,n})| & |K(Z_{+,n})| & h^{2,1}(Z_{+,n})  \\
    \hline
  0 &  2 &  0 &  112  \\
  1 &  2 &  0 &  112  \\
  2 &  6 & 0 & 128    \\
  3 &  8 & 0 & 154  \\
  4 &  9 & 0 & 182    \\
  5 & 10 & 1 & 211    \\
  6 & 10 & 0 & 240   
 \end{array}\, .
\end{equation}

As we have engineered our TCS $G_2$ models such that $Z_{+,n}$ carries all the bundle data, the $G_2$-manifolds $M_n$ dual to heterotic models on $X_{\rm het}$ with a distribution of instantons \eqref{eq:instdistro} are constructed as a twisted connected sum of the building blocks $Z_{+,n}$ with the same $Z_-$, \eqref{eq:bbe8e8}, throughout. In particular, the hyper-K\"ahler rotations which are used in the matching descend from the same lattice autmorphism for each of those models. Viewing the building blocks $Z_{+,n}$ as K3 fibrations, the resolution of the singularities \eqref{eq:nhcbb} results in fibers with a lattice $N = U \oplus G$. The identification of lattices we want to use to define a hyper-K\"ahler rotation then results in 
\begin{equation}
 N_+ \cap N_- = G \, .
\end{equation}
This means that the matching \eqref{eq:DM} forces the K\"ahler forms in all of the K3-fibers for both building blocks to integrate to zero over the cycles contained in the lattice $G$. In other words, there is an ADE singularity of type $G$ in every K3-fiber of the building blocks, and hence over every point of the $S^3$ base of the glued $G_2$-manifold. Although there is no rigorous mathematical argument at present that these singularities will persist when the metric of the glued manifold is perturbed such that it becomes Ricci-flat, we conjecture based on the duality to the heterotic string (and also F-theory) that this is indeed the case. 

To find the spectra of M-theory on the resulting $G_2$-manifolds, we proceed as follows. Using mirror symmetry for $G_2$-manifolds, the formulae \eqref{eq:bettinumbersTCS} and \eqref{eq:b2b3sumruleorthgluing} correctly reproduce the rank of the gauge group and the number of uncharged chiral multiplets even in the presence of the singularities we encounter. The reason is that we can confidently determine $b_2(J)$ by couting the rank of the gauge group and that we have a smooth mirror $J^\vee$ with $N_+ \cap N_- = 0$. We can hence find $b_3$ via $b_2(J^\vee) + b_3(J^\vee) = b_2(J) + b_3(J)$. It would be nice to have a rigorous mathematical argument or find the cohomology theory which reproduces these numbers in the presence of singularities. In the present case, the number $n_v$ of $U(1)$ vectors, neglecting the Cartan of the non-abelian gauge group $G$, the non-abelian gauge group $G$ and the number of chiral multiplets can then be computed from \eqref{eq:databbn} and \eqref{eq:databb-} to be 
\begin{equation}\label{eq:specG2}
 \begin{array}{c|ccccc}
  n & G & n_v  & n_c \\
    \hline
  0 & - & 12& 299\\
  1 & - & 12& 299 \\
  2 & SO(8)  & 12 & 327 \\
  3 & E_6  & 12 & 377 \\
  4 & E_7  & 12 & 432 \\
  5 & E_8  & 13 & 490 \\
  6 & E_8  & 12 & 547
 \end{array}
\end{equation}
We will recover this spectrum from the perspective of F-theory in the next section.

\section{F-Theory Lift}\label{sect:F-theory}

To lift this model to F-Theory, we simply note that the heterotic three-fold $X_{\rm het}$ is an elliptic fibration over $dP_9$ and replace the elliptic curve $\mathbb{E}$ by an elliptic K3-surface, with fiber $\mathbb{E}$. This means we consider K3-fibered Calabi-Yau fourfolds over $dP_9$ as potential F-Theory duals. Generic fourfolds $X_F$ of this type were considered in \cite{Donagi:1996yf}. 

\subsection{The Geometry}

To realize a K3-fibration over $dP_9$, we construct both the $dP_9$ and the K3-surface as a Weierstrass model. The different instanton distributions \eqref{eq:instdistro} are then realized by fibering the base of the K3-surface over the base of the $dP_9$ in an approriate way. This means we realize the elliptic fourfolds $X_{F,n}$ as complete intersections in ambient spaces with weight systems
\begin{equation}\label{eq:weightsftheory}
 \begin{array}{rrrrrrrrrr|c|rr}
 x & y & w & \hat{x} & \hat{y} & \hat{w} & z_1 & z_2 & \hat{z}_1 & \hat{z}_2 & \Sigma \hbox{ of degrees}& W & \hat{W} \\\hline
 2 & 3 & 1 & 0 & 0 & 0 & 0 & 0 & 0 & 0 & 6 & 6 & 0\\
 0 & 0 & 0 & 2 & 3 & 1 & 0 & 0 & 0 & 0 & 6 & 0 & 6\\
 4 & 6 & 0 & 0 & 0 & 0 & 1 & 1 & 0 & 0 & 12 & 12& 0\\
 2 +2n& 3 +3 n& 0 & 2 & 3 & 0 & 0 & n & 1 & 1 & 12 +n  & 6+n& 6+n
 \end{array}
\end{equation}
The first equation has the form
\begin{equation}\label{eq:hatW}
\hat{W} = - \hat{y}^2 + \hat{x}^3 + \hat{x} \hat{w}^4 \hat{f}_{4}(\hat{z}) +  \hat{w}^6 \hat{g}_{6}(\hat{z}) = 0\,,
\end{equation}
where $\hat{f}$ and $\hat{g}$ are of the indicated degree in $\hat{z}_1$ and $\hat{z}_2$. If we project out the unhatted coordinates, this realizes a $dP_9$ as an elliptic fibration over the $\widehat{\P}^1$ with homogeneous coordinates $\hat{z}_1:\hat{z}_2$. Together with the $\P^1$ with coordinates $[z_1:z_2]$ it forms the base of our elliptic fourfold, which is now described by intersecting the above with
\begin{equation}\label{eq:W}
W =  - y^2 + x^3 + x w^4 f_{8,4+4n}(z,\hat{z}) + w^6 g_{12,6+6n}(z,\hat{z}) = 0\, .
\end{equation}
Here, $f_{8,4}(z,\hat{z})$ and $g_{12,6}(z,\hat{z})$ are homogeneous of the indicated weights. 

\begin{figure}
\begin{center}
  \includegraphics[height=6cm]{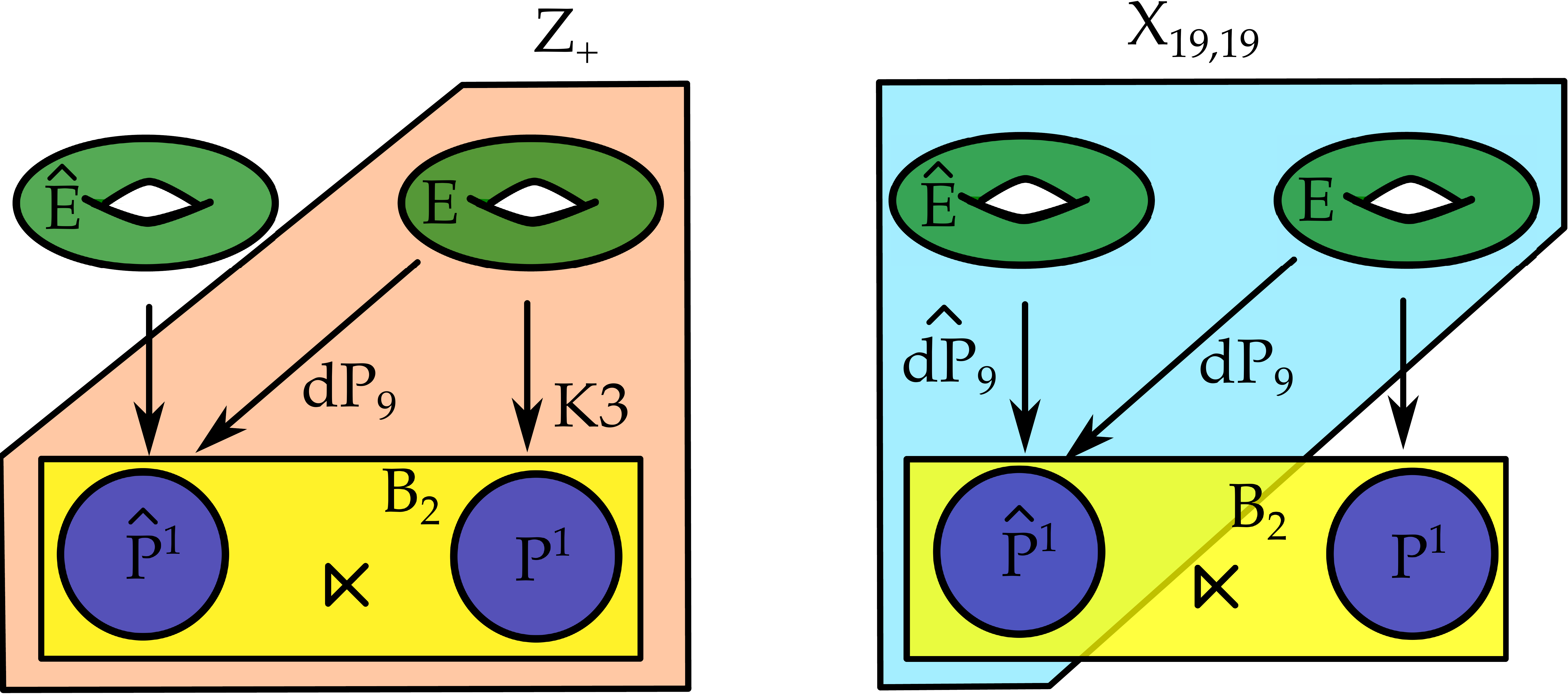}
 \caption{\label{fig:XF} A sketch of the geometry $X_{F,n}$. The two elliptic curves $\widehat{\mathbb{E}}$ and $\mathbb{E}$ are shown as well as the non-trivial way they are fibered over the base $B_2=\widehat{\mathbb{P}}^1 \ltimes \mathbb{P}^1$. The fibration result in the $dP_9$s, as well as the K3, as indicated by the boxes. For the general construction in section \ref{sec:GeneralStory}, the base $B_2$ is required to only have a projection to $\widehat{\mathbb{P}}^1$ and its fiber $\P^1$ is allowed to have degenerations into multiple components over points in the base 
 $\widehat{\mathbb{P}}^1$. }
\end{center}
\end{figure}

Note that this realization is very similar to our construction of the building blocks $Z_{+,n}$, \eqref{eq:weightsysfnbb}. In fact, $X_{F,n}$ is realized as the fiber product 
\be\label{eq:fibreproduct}
X_{F,n} = Z_{+,n}\times_{\widehat{\P}^1} dP_9\,,
\ee
where the common $\widehat{\P}^1$ is the one with coordinates $[\hat{z}_1:\hat{z}_2]$ \footnote{We thank Dave Morrison for pointing this out.}. A sketch of the geometry is shown in figure \ref{fig:XF}. 
It is hence not surprising that the complete intersections presented above are also singular when $n\geq 2$, and we can repeat the same table as shown in \eqref{eq:nhcbb} for the appearing singularities. The corresponding gauge groups are geometrically non-Higgsable in the sense of \cite{Grassi:2014zxa, Morrison:2014lca}.

The above presentation fits with the framework of \cite{Batyrev:1994pg}, i.e. the complete intersection $W =\hat{W} = 0$ is associated with a nef partition of $-K$ of the ambient space and there is an associated reflexive polytope $\Delta^\circ$ which can directly be found from \eqref{eq:weightsftheory} in the case $n\leq 1$. This is not the case for $n \geq 2$, which corresponds to the fact that these cases require resolution. These resolutions again have an elegant construction in terms of polytopes, see Appendix \ref{app:4folds} for the details. After the resolution is performed with these methods, we can easily compute the Hodge numbers and hence the spectra of the F-theory compactification.

In the case $n=5$ there are loci over which the vanishing orders of $f,g,\Delta$ are  $\hbox{ord} (f,g,\Delta)=(4,6,12)$. This signals the presence of tensionless strings in F-Theory from wrapped D3 branes on collapsed cycles in the base of singular limit of the fourfold $X_{F,5}$. These non-minimal points can be blown up, such that the F-Theory compactification is captured by an effective field theory. This is the fourfold we are working with. 
In this case, the Hodge numbers of the base $B_3$ of the elliptic fibration are hence not $h^{1,1}=11$ and $h^{2,1}=0$ as in all the other cases, but $h^{1,1}=12$ and $h^{2,1}=1$. 

In line with our general picture \eqref{eq:fibreproduct}, this indicates the presence of tensionless strings for the M-Theory compactification on the $G_2$-manifold $J$ glued from $Z_{+,5}$ and $Z_{-,5}$. In fact, the Donaldson matching forces to collapse the cycles in the K3 fibres of $Z_{+,5}$ corresponding to $E_8$, as we have discussed above. In light of the existence of tensionless strings in the dual F-Theory model, we conjecture that this will force the collapse of a coassociative four-cycle which arises as a component of a reducible K3 fibre. The tensionless strings then arise from wrapped M5-branes wrapped on this coassociative cycle. The 2d theories on the string are 2d (0,2) supersymmetric \cite{Gadde:2013sca, Assel:2016lad}.  

For the comparision of the spectra, we perform the corresponding blowup at the level of the building block $Z_{+,5}$, which can be done by trading the top $\Diamond^\circ_5$ for $\Diamond^{\circ *}_5$, see \eqref{eq:diamondsareforever}. This does not change any of the relevant topological data for $Z_{+,5}$. From the point of view of the elliptic fibration, this can be understood as a flop which maps a non-flat fibration into a flat one, where the base in turn has been blown up. 

\subsection{$G_4$-Flux}
\label{sec:G4Flux}

An F-theory compactification to 4d is specified by the geometry of the elliptic fibration as well as background $G_4$-flux. 
The tadpole cancellation condition in F-theory on an elliptic Calabi-Yau four-fold is
\be\label{D3Tad}
{\chi(X_F)\over 24} = N_{D3} + {1\over 2} \int_{X_F} G_4 \wedge G_4 \,,
\ee
where $G_4$ is the four-form flux and $N_{D3}$ the number of D3-branes. The number of D3-branes therefore depends on the flux turned on in the background. 
Furthermore, we need to impose flux quantization \cite{Witten:1996md}
\be
G_4 + {c_2 (X_F) \over 2} \in H^4 (X_F, \mathbb{Z})\,,
\ee
where $X_F$ is the smooth, resolved elliptic Calabi-Yau four-fold. 
If $c_2$ is odd, the minimal flux that has to be switched on is therefore $c_2 (X_F)/2$. Chirality is computed by integrating the flux over matter surfaces, however there are no codimension two enhancements whereby there are no non-trivial matter loci in all our models. 

By direct computation of the second Chern classes in the models that we consider below, it in fact follows that $c_2 (X_{F, n})  = 0 \mod 2$, and thus the minimal flux can be 0. In section \ref{sec:GeneralStory} we shall prove quite generally that $c_2$ is even based on the structure of $X_{F}$. 

In fact this can also be observed by considering the fluxes obtained in 
\cite{Marsano:2011hv, Kuntzler:2012bu, Krause:2012yh} from explicit resolutions and the so-called spectral divisor, which implies that the flux, which is both properly quantized and does not break the gauge group, will be proportional to the so-called matter-surface, i.e. the rational curves in the singular fiber above the codimension two matter loci in the base. Again in the case of $X_{F, n}$, this flux vanishes by the absence of the matter loci.

\subsection{Spectrum}

As $G_4=0$, the table \eqref{eq:nhcbb} directly tells us the non-abelian gauge group of the effective 4d field theory, which matches the non-abelian gauge groups on the M-Theory side. As furthermore $\chi(X_{F,n}) = 288$ in all cases, we need to include $12$ D3-branes in each model, due to tadpole cancellation. These correspond to the $12$ NS5-branes wrapped on $\mathbb{E}$ in the dual heterotic models and together contribute $12$ $U(1)$ vector- and $36$ chiral multiplets. Finally, there are no matter curves, so that the rest of the spectrum is given purely in terms of geometry. The number of chiral multiplets in each model is hence
\begin{equation}\label{eq:fthchiralmult}
\ba
 n_c =&  h^{1,1}(B) +  h^{2,1}(X_{F}) + h^{3,1}(X_{F}) - h^{2,1}(B) + 3 N_{D3}\cr 
=&  h^{1,1}(B) + h^{2,1}(X_{F,n}) + h^{3,1}(X_{F,n}) - h^{2,1}(B) + 36
\ea
\end{equation}
and the number of $U(1)$ vectors neglecting the non-abelian gauge group content is
\begin{equation}\label{eq:ftheoryvmult}
\ba
 n_v =&  h^{1,1}(X_{F}) - h^{1,1}(B) - 1 + h^{2,1}(B) + N_{D3} - \mbox{rank}(G) \cr 
 =& h^{1,1}(X_{F,n}) - h^{1,1}(B) - 1 + h^{2,1}(B) + 12 - \mbox{rank}(G) \, .
\ea
\end{equation}
Together with the Hodge numbers of the $X_{F,n}$ and the Hodge numbers of their base space we hence find 
\begin{equation}
 \begin{array}{c|ccccccc}
  n & G & h^{1,1} (X_{F,n})& h^{2,1}(X_{F,n}) & h^{3,1} (X_{F,n})& \chi (X_{F,n})& n_c & n_v \\
  \hline
  0 & -&12 & 112 & 140 & 288 & 299 & 12 \\
  1 & -&12 & 112 & 140 & 288 & 299 & 12 \\
  2 & SO(8)&16 & 128 & 152 & 288 & 327 & 12 \\
  3 & E_6&18 & 154 & 176 & 288 & 377 & 12 \\ 
  4 & E_7&19 & 182 & 203 & 288 & 432 & 12 \\
  5 & E_8&21 & 212 & 231 & 288 & 490 & 13 \\
  6 & E_8&20 & 240 & 260 & 288 & 547 & 12 \, 
 \end{array}
\end{equation}
which precisely matches with the spectrum found from the $G_2$-manifolds on the M-Theory side \eqref{eq:specG2}! Besides a beautiful matching of the moduli counting, this also provides a dual description of non-Higgsable gauge groups in M-Theory. 
Finally let us comment on the non-Higgsability: the F-theory four-folds realize geometrically non-Higgsable gauge theories. 
In models where there are non-trivial matter curves in the base of the F-theory four-fold, i.e. codimension two enhancements of the discriminant, we would expect to have vector-like matter even in the absence of flux, unless the matter curves are rational. Such vector-like matter could potentially Higgs the gauge group. In the present context, there are no matter loci, and thus the geometric non-Higgsability implies non-Higgsability of the 4d effective theory.


\section{Generalization}
\label{sec:GeneralStory}

In this section we generalize the construction made in earlier sections of this paper to include more general non-abelian gauge groups. The idea is to replace the building block $Z_+$ to have a more general base $B_2$, as shown in  figure \ref{fig:XF}.

\subsection{General Construction and non-Abelian Gauge Groups}

To motivate our generalization, consider the decomposition of the Calabi-Yau fourfolds $X_F$ constructed above.  We may view the manifolds $X_F$ as elliptic fibrations over the building blocks $Z_+$, where the elliptic curve $\widehat{\mathbb{E}}$ of this fibration only varies non-trivially over the base $\widehat{\P}^1$ of the K3-fibered threefold $Z_+$. Note that the elliptic curve, whose complex structure is identified with the axio-dilaton of F-theory, is $\mathbb{E}$.

This immediately leads us to conjecture that such a duality holds for any algebraic two-dimensional base $B_2$, which has a projection to $\widehat{\P}^1$ and the generic fiber of which is a $\P^1$. In contrast to the simple examples of $B_2 =\mathbb{F}_n$ considered before, the fiber $\P^1$ may undergo degeneration into a bouquet of $\P^1$s over points in $\widehat{\P}^1$. From such a two-dimensional base $B_2$, the building block $Z_+$ is then constructed as a Weierstrass elliptic fibration over $B_2$ such that $c_1(Z_+) = [\hat{H}]$, where $\hat{H}$ is the hyperplane class of $\widehat{\P}^1$. As $B_2$ has a projection to $\widehat{\P}^1$, the same projection gives us a K3-fibration on $Z_+$ with generic fiber $S_+$. We now have $c_1(Z_+) = [\hat{H}] = [S_+]$ as expected for a building block. From this, the F-Theory geometry $X_F$ is constructed as an elliptic fibration over $Z_+$ with fiber $\widehat{\mathbb{E}}$, such that $\widehat{\mathbb{E}}$ degenerates over $12$ points of $\widehat{\P}^1$ but has no non-trivial dependence on $S_+$. This ensures that $X_F$ is a Calabi-Yau fourfold. The geometry is again depicted in figure \ref{fig:XF}, with the suitable modifications of the base $B_2$. Note that F-Theory compactifications in which the vanishing degrees of $(f,g,\Delta)$ are greater or equal to $(4,6,12)$ give rise to tensionless strings. This implies that tensionsless strings also appear in M-Theory on TCS $G_2$-manifolds with the corresponding $Z_+$ and $Z_-$. Here, they arise from M5-branes wrapped on collapsed coassociative cycles. Again, the theories can be studied along the lines of \cite{Gadde:2013sca, Assel:2016lad}.

The $G_2$-manifold $J$, which gives the M-theory dual of F-Theory on $X_F$ is constructed as a twisted connected sum of $Z_+$ with $Z_-$, where we take the same $Z_-$ as before, namely the one constructed via the tops \eqref{eq:bbe8e8}, for every $J$. As before, the matching between $Z_+$ and $Z_-$ is such that we identify the two distinguished $E_8$ lattices in $H^2(S_+,\mathbb{Z})$ and $H^2(S_-,\mathbb{Z})$. These lattices are distinguished as we assume that both $Z_\pm$ are constructed as Weierstrass models. In $Z_-$ the $E_8 \oplus E_8$ lattice corresponds to the two $II^*$ fibers. As $Z_+$ is also a Weierstrass model, there is a point in the moduli space of the fiber (as an algebraic family) in which it develops two $II^*$ fibers. This defines the distinguished $E_8 \oplus E_8$ for $Z_+$. Note that this implies that the $G_2$-manifold $J$ will be forced to have ADE singularities whenever there are $(-2)$-curves in the distinguished $E_8\oplus E_8$ lattice, which sit in $N$. This happens whenever all of the $S_+$ over $\widehat{\P}^1$ have elliptic fibrations with reducible elliptic fibers.

The elliptic fibration 
\begin{equation}
 \mathbb{E} \hookrightarrow X_F \rightarrow B_3 \,,\qquad \hbox{where }\qquad \widehat{\mathbb{E}} \hookrightarrow \ B_3  \ \rightarrow B_2 \,,
\end{equation}
of $X_F$, which gives a dual description of M-Theory on $J$, is the same as the elliptic fibration on $Z_+$ and constant over $\widehat{\mathbb{E}}$.
An elliptic fibration with reducible fibers on $S_+$ (present over all of $\widehat{\P}^1$), which gives us ADE singularities on $J$ as discussed above, hence gives rise to the corresponding ADE gauge groups on the F-Theory side. 
Futhermore, the matter loci occur at points on $Z_+$. In F-theory, these are matter curves, (i.e. codimension two singular loci in the base) of $X_F$, each of which  is a copy of $\widehat{\mathbb{E}}$. 
In the M-theory geometry $J$ these loci are circles $\mathbb{S}^1_{e+}$, and their number and type agrees with the matter curves in F-theory. 

We conjecture that the singular TCS $G_2$-manifolds which underlie the M-theory compactifications exist and that they give rise to the same gauge theories on the F-Theory side. Note that while the ADE singularities arise in the limit in which $\mathbb{E}$ is collapsed on the F-theory side, it is the Donaldson matching with $Z_-$ which forces to collapse the relevant cycle on the M-Theory side. 

To provide evidence for this conjecture, we will prove below that the spectra of uncharged particles and the number of $U(1)$ factors of the gauge group agree on both sides for any base geometries $B_2$ with the properties that $B_2$ is a smooth algebraic surface and has a projection to $\widehat{\P}^1$ with generic fiber $\P^1$. Our discussion does not rely on a specific construction of the building block $Z_+$ or the Calabi-Yau fourfold $X_F$. We will illustrate this method by constructing a few more examples below. 

\subsection{Comparison of Spectra}

To substantiate the construction of dual pairs of $G_2$-manifolds $J$ and F-theory Calabi-Yau four-folds $X_F$, we now compute the spectra on both side, and check some global consistency conditions. 

Let us first express the spectrum on the M-theory side in terms of the Hodge numbers of the building block $Z_+$. From \eqref{eq:b2b3sumruleorthgluing} and \eqref{eq:databb-} we have
\begin{equation}\label{eq:specM}
 \begin{aligned}
 n_v & = |N_+ \cap N_-| + |K(Z_+)| + |K(Z_-)| &=& |G| + |K(Z_+)| + 12 \\
 n_v + n_c  &= 23 + 2(h^{2,1}(Z_+) + h^{2,1}(Z_-) + 2 (K(Z_+) + K(Z_-)) &=& 2\left(h^{2,1}(Z_+) + |K(Z_+)| \right) + 87 \, ,
 \end{aligned}
\end{equation}
where $|G|$ is the rank of the non-abelian gauge group $G$, which may contain $U(1)$ factors from extra sections of the elliptic fibration of $Z_+$ by $\mathbb{E}$.   
In our construction the Donaldson matching is such that 
\be
N_+ = U \oplus G\qquad \hbox{and}\qquad
N_+ \cap N_- = G \,,
\ee
and such sections realizing $U(1)$ symmetries enter in the same way as non-abelian factors. The lattice $G$ has a root sublattice $G_{\rm root}$, generated by lattice vectors of length $-2$, which determines the non-abelian gauge group. The number of $U(1)$s in $G$ is then given by $\rk(G)-\rk(G_{\rm root})$. To compare this to the spectrum on the F-theory side, \eqref{eq:fthchiralmult} and \eqref{eq:ftheoryvmult}, we need to relate the Hodge numbers of $Z_+$ with those of $X_F$ and $B_3$. Note that in our notation, $n_c$ counts the number of chiral multiplets neutral under $G$. 

The fourfold $X_F$ constructed above has two elliptic fibrations. Although the fibration by $\mathbb{E}$ defines the F-theory elliptic fibration of $X_F$, i.e. the complex structure is identified with the axio-dilaton, 
$X_F$, is also elliptically fibered over $Z_+$ with fiber $\widehat{\mathbb{E}}$
\begin{equation}\label{eq:ehatfibxf}
 \widehat{\mathbb{E}} \hookrightarrow X_F \rightarrow Z_+ \, .
\end{equation}
A generic elliptic Calabi-Yau $X_F$ with fiber $\widehat{\mathbb{E}}$ over $Z_+$ only has $12$ singular fibers over $\widehat{\P}^1$ and no reducible fibers. This way of looking at $X_F$ is key to showing the equivalence of the spectra. 

We start by showing that the second Chern class of $X_F$ is always even, so that we do not have to include any 
$G_4$-flux in order to have a well-defined model, as discussed in section \ref{sec:G4Flux}. Using adjunction and $\hat{\sigma}\cdot (\hat{\sigma} +\hat{H})= 0$ for the section $\hat{\sigma}$ of the elliptic fibration by $\widehat{\mathbb{E}}$ and the hyperplane class $\hat{H}$ of $\widehat{\P}^1$ we find
\begin{equation}
\begin{aligned}
 c_2(X_F) &= 11 \hat{\sigma}^2 + 23 \hat{H} \cdot \hat{\sigma} + c_2(Z_+) \\
 & = 12 \hat{H} \cdot \hat{\sigma} + c_2(Z_+) \, ,
\end{aligned}
\end{equation}
so that $c_2(X_F)$ is even if and only if $c_2(Z_+)$ is even. This, however, is guaranteed for any algebraic threefold, see  Lemma 5.10 of \cite{MR3109862} for a discussion in the context of TCS building blocks. In fact, this argument is a special case of the one given in \cite{Collinucci:2010gz} for elliptic fourfolds without reducible elliptic fibers in codimension one in the base. Thus in summary, for any such model 
\be
{c_2 (X_F)\over 2} \in H^4 (X_F, \mathbb{Z}) \,.
\ee
Next, we compute the Euler characteristic of $X_F$ using the elliptic fibration by $\widehat{\mathbb{E}}$ again. To compute the Euler characteristic, we consider the $12$ points $\hat{p}_i$ in $\widehat{\P}^1$ over which the fiber  $\widehat{\mathbb{E}}$ is singular,  separately from $\widehat{\P}^1$ with those twelve points excised and sum the results. We may further decompose $\hat{\P}\setminus\{\hat{p}_i\}$ into simply connected patches, each of which is topologically $\widehat{\mathbb{E}} \times U$, for a disc $U$, and hence contributes zero to the Euler characteristic. This makes it clear that only the $12$ points $\hat{p}_i$ give a non-zero contribution. In a generic situation, fixing any $\hat{p}_i$ on $Z_+$ gives a copy of a generic (i.e. irreducible and smooth) K3-fiber of $Z_+$, times a  singular $\widehat{\mathbb{E}}$, which has $\chi =1$. We can hence compute
\begin{equation}
 \chi(X_F) = 12  \times 1 \times \chi(K3) = 288 \,, 
\end{equation}
for any fourfold $X_F$ of the form \eqref{eq:ehatfibxf}. Note that, together with the absence of $G_4$-flux, the D3-brane tadpole (\ref{D3Tad}) implies that every such model requires $\chi (X_F) /24 = 12$ D3-branes, which contribute $36$ chiral multiplets and $12$ vector multiplets in four dimensions. The number of vector multiplets matches with the contribution of $|K(Z_-)|=12$ from $Z_-$ via \eqref{eq:b2b3sumruleorthgluing}, which also stays invariant for all of the models we consider. 

A similar argument shows that the base $B_3$ of the elliptic fibration with fiber $\mathbb{E}$ always has Euler characteristic $24$. By construction, $B_3$ is also elliptic with fiber $\widehat{\mathbb{E}}$, which degenerates over $12$ points in $\widehat{\P}^1$, i.e. over $12$ copies of  $\P^1$ in $B_2$. Hence
\begin{equation}\label{eq:chibase}
 \chi(B_3) = 12 \times  \chi(\P^1) = 24 \, .
\end{equation}

The fact that $X_F$ is an elliptic fibration over $Z_+$ without singular fibers but with $8$ distinct extra sections (in addition to the zero-section), which allows us to directly conclude that
\begin{equation}\label{eq:h11zh11x}
 h^{1,1}(X_F) = h^{1,1}(Z_+) + 9 \,,
\end{equation}
where one contribution is due to the fiber class $\widehat{\mathbb{E}}$.

To find $h^{2,1}(X_F)$, we need to work a little bit harder. For this computation, it is advantageous to view $X_F$ as a fibration of $\widehat{\mathbb{E}} \times S_+$ over $\widehat{\P}^1$. Conveniently, both $\widehat{\mathbb{E}}$ and $S_+$ degenerate over different points in $\widehat{\P}^1$. This allows us to cut $\widehat{\P}^1$ into two overlapping open discs $X_{\widehat{\mathbb{E}}}$ and $X_{Z_+}$ such that $X_{\widehat{\mathbb{E}}} = (dP_9 \setminus \widehat{\mathbb{E}}) \times S_+$ and $X_{Z_+} = (Z_+\setminus S_+) \times \widehat{\mathbb{E}}$, i.e. only one of the two fibers varies non-trivially in each of the two halves of $\widehat{\P}^1$. This allows to use the Mayer-Vietoris sequence for $X_F = X_{\widehat{\mathbb{E}}} \cup X_{Z_+}$ with $X_{\widehat{\mathbb{E}}} \cap X_{Z_+}$ retracting to $\widehat{\mathbb{E}} \times S_+ \times \mathbb{S}^1$. With 
\begin{equation}
\gamma^k : H^{k}( X_{\widehat{\mathbb{E}}} ) \oplus H^{k}( X_{Z_+} ) \rightarrow H^k(X_{\widehat{\mathbb{E}}} \cap X_{Z_+}) \, 
\end{equation}
we then have
\begin{equation}
H^k(X_F) = \mbox{ker}(\gamma^k ) \oplus \mbox{coker}(\gamma^{k-1}) \, .
\end{equation}
This allows us to rederive \eqref{eq:h11zh11x}, but also gives 
\begin{equation}
b_3(X_F) = \mbox{ker}(\gamma^3 ) =  2\left( |K(Z_+)| + h^{1,1}(Z_+) \right)
\end{equation}
as $\mbox{coker}(\gamma^{2}) = 0$. As $h^{3,0}(X_F) = 0$ it follows that
\begin{equation}
h^{2,1}(X_F) = |K(Z_+)| + h^{1,1}(Z_+) \, .
\end{equation}
We can now exploit $\chi(X_F) = 288$ together with
\begin{equation}
h^{2,2}(X_F) = 2\left( 22+2h^{1,1}(X_F) +2h^{3,1}(X_F) - h^{2,1}(X_F) \right)
\end{equation}
to find
\begin{equation}
h^{3,1}(X_F) = h^{2,1}(Z_+) - h^{1,1}(Z_+) + |K(Z_+)|  + 31 \, . 
\end{equation}

For the base $B_3$ of the elliptic fibration of $X_F$ with fiber $\mathbb{E}$ we find
\begin{equation}
h^{1,1}(B_3) = h^{1,1}(B_2) + 9  = |K(Z_+)| + 11  \, , 
\end{equation}
by exploiting its elliptic fibration with fiber $\widehat{\mathbb{E}}$ and base $B_2$, as well as $h^{1,1}(B_2) = |K(Z_+)| + 2$. Furthermore, as $X_F$ is an elliptic fibration over $B_3$ and Calabi-Yau, $h^{i,0}(X_F) =0$ for $i =1,2,3$, it follows that  $h^{3,0}(B_3) = 0$, as any non-trivial such class of the base would pull-back to $X_F$. 

We can now use adjunction for the Weierstrass model with elliptic fiber $\widehat{\mathbb{E}}$ to find that 
\begin{equation}
 \chi_0(B_3) =  \frac{1}{24} \int_{B_3}  c_1(B_3) c_2(B_3) = \frac{1}{24} \int_{B_2} 12 c_1(B_2) \cdot \hat{H} =\frac{1}{24} \int_{B_3} c_3(B_3) = \chi(B_3) = 1 \, .
\end{equation}
As
\begin{equation}
 \chi_0(B_3) = 1 - h^{1,0}(B_3)  + h^{2,0}(B_3)  -h^{3,0}(B_3)  = 1
\end{equation}
and $h^{3,0}(B_3) =0$ we find $h^{1,0}(B_3) -h^{2,0}(B_3) =0$, so that finally
\begin{equation}
h^{2,1}(B_3) = |K(Z_+)| 
\end{equation}
follows from $\chi(B_3) = 24$. 

We are now ready to compute the spectrum on the F-Theory side in terms of $Z_+$. First we compute the rank of the total gauge group
\begin{equation}\label{eq:fnv}
\begin{aligned}
n_v & = h^{1,1}(X_{F}) - h^{1,1}(B_3) - 1 + h^{2,1}(B_3) + 12 \\
& = h^{1,1}(Z_+) + 9 - |K| - 11 - 1 + |K| + 12 \\
& = h^{1,1}(Z_+) + 9 \\
& = |G| + |K(Z_+)| + 12 \cr 
& = h^{1,1} (X_F) \,,
\end{aligned}
\end{equation}
where we have exploited the elliptic fibration on $Z_+$ with fiber $\mathbb{E}$ and $h^{1,1}(B_2) = |K(Z_+)| + 2$ in the last line. The summand $|G|$ counts the rank of the non-abelian gauge group from reducible fibers of $\mathbb{E}$ together with extra sections of the elliptic fibration by $\mathbb{E}$. Note that the number of sections of the elliptic fibration of $X_F$ by $\mathbb{E}$ is equal to the number of sections of the elliptic fibration of $Z_+$ by $\mathbb{E}$, as $\mathbb{E}$ is constant over $\widehat{\mathbb{E}}$.
Next we check \
\begin{equation}\label{eq:fnvnc}
\begin{aligned}
 n_v + n_c &= h^{1,1}(X_{F}) + h^{2,1}(X_{F}) + h^{3,1}(X_{F})  +  47 \\
 &=  9 + h^{2,1}(Z_+) + |K(Z_+)| +  h^{2,1}(Z_+)  + |K(Z_+)|  + 31 + 47 \\
 & = 2\left( h^{2,1}(Z_+) + |K(Z_+))| \right) + 87 \, ,
\end{aligned}
\end{equation}
where $n_c$ again counts the number of chiral multiplets neutral under $G$. Comparing \eqref{eq:fnv} and \eqref{eq:fnvnc} with \eqref{eq:specM}, we find a perfect agreement of the spectra between the dual compactifications of M-theory on the $G_2$-manifold $J$ and F-theory on the elliptic fourfold $X_F$. Note that both theories have the gauge group
\begin{equation}
 G \times U(1)^{|K(Z_+)| + 12} \, ,
\end{equation}
where $G$ contains non-abelian factors from $G_{\rm root}$ and $\rk(G)-\rk(G_{\rm root})$ abelian gauge factors.

\subsection{Some TCS $G_2$-Manifolds for Non-abelian Theories}

Let us bring the general proof given above to life and consider a few more examples of F-theory/M-theory dual pairs. In particular, let us consider situations in which there is a Higgsable non-abelian gauge group (we consider split and non-split models). 

\subsubsection{Dual Models with $G = SU(2)$}

In this model we consider an elliptic building block $Z_+$ with $N_+ = U \oplus A_1$. Such a model is found from the dual pair of tops with vertices
\begin{equation}
 \Diamond^\circ = \left(\begin{array}{rrrrrr}
 x& y & z_\delta & z_2 & \hat{z}_1 & z_1 \cr
-1 & 0 & 1 & 2 & 2 & 2 \\
0 & -1 & 2 & 3 & 3 & 3 \\
0 & 0 & 1 & -1 & 0 & 1 \\
0 & 0 & 0 & 0 & 1 & 0
\end{array}\right) \, ,\hspace{1cm}
\Diamond = \left(\begin{array}{rrrrrrrr}
-2 & -1 & 1 & -1 & 1 & 1 & 1 & 1 \\
1 & 1 & 1 & 1 & 1 & 1 & 1 & -1 \\
0 & -2 & 6 & -2 & 6 & -4 & -4 & 0 \\
0 & -2 & 0 & 0 & -6 & 0 & -6 & 0
\end{array}\right)\, .
\end{equation}
The hypersurface (\ref{HypSurf}) is obtained by first finding all the lattice points $\nu_i$ on $\Diamond^\circ$. Together  with $\nu_0$ these generate the rays of the fan of the toric variety $\mathbb{P}_\Sigma$ in which the hypersurface $Z_+$ is embedded. In the present example, this gives rise to the blowup of the $\mathbb{P}_{1,2,3}$ bundle over the base $B_2= \widehat{\mathbb{P}}^1 \times \mathbb{P}^1$ resolving the Kodaira $I_2$ fiber. The correspondence between the homogeneous coordinates of $\widehat{\mathbb{P}}^1 \times \mathbb{P}^1$ and the rays of $\Sigma$ is as follows:
\be
\begin{array}{c|l} 
 \widehat{\mathbb{P}}^1 \times \mathbb{P}^1 & \hbox{Ray generator of } \mathbb{P}_\Sigma \cr \hline 
z_1 &(  2, 3, 1, 0 )\cr 
z_2 & (2,3,-1, 0) \cr 
\hat{z}_1 & (2,3,0,1) \cr 
\hat{z}_2 & (0,0,0,-1) = \nu_0
\end{array}
\ee
Note that the corresponding Tate model does not have a trivial canonical class, due to the distinguished role of $\nu_0$. This is responsible for $c_1(Z_+)= [\hat{z}_2] = \hat{H}$.

The building block $Z_+$ is described by a Weierstrass model  
\begin{equation}
 y^2 + b_1 x y w+ b_3 z_1 y w^2  =z_\delta x^3 + x^2  w^2 b_2 + x w^4  z_1 b_4 +   w^6 z_1^2 b_6  \,, 
\end{equation}
where $z_\delta$ is the section corresponding to the resolution divisor blowing up $x=y=z_1=0$. 
The $b_i$ are sections of $K_{\mathbb{P}^1}^{-i} \otimes K_{\widehat{\P}^1}^{-i/2} \otimes \mathcal{O}(H^{-n_i})$ where $\vec{n}= (0,0,1,1,2)$ are the vanishing orders of the $I_2$ Tate model.  
The $I_2$ fiber degenerates further over 
\be
\Delta = (b_1^2 + 4 b2)^2  \left((b_6 b_1^2-b_4 \left(b_1 b_3+b_4\right)+b_2
   \left(b_3^2+4 b_6\right)\right) =0 \,.
\ee
These degeneration loci are present over points of $Z_+$, so that the matter on the M-theory side is localized on circles $\mathbb{S}_e^1$, see (\ref{sec:TCSG2}). 
On the F-theory side, the same degenerations of the elliptic fibration happen over 
 matter curves that are tori (the elliptic curves $\widehat{\mathbb{E}}$), and thus in the absence of $G_4$-flux do not generate any chiral matter. 

Finally, let us compare the uncharged spectra of the dual models: 
From \eqref{eq:bbtopology} it follows that
\begin{equation}
h^{1,1}(Z_+) =  4 \hspace{1cm} h^{2,1}(Z_+) =  98 \hspace{1cm} |N(Z_+)| = 3\hspace{1cm} |K(Z_+)| = 0 \, ,
\end{equation}
so that the Betti numbers of the $G_2$-manifold $J$ are 
\begin{equation}
 b_2 = 13 \, , \hspace{1cm} b_3 = 270  \, .
\end{equation}
The dual F-Theory model is compactified on a fourfold $X_F$ with Hodge numbers
\begin{equation}
h^{1,1}(X_F)=13 \, ,\hspace{1cm} h^{2,1}(X_F)=98 \, ,\hspace{1cm}  h^{3,1}(X_F)=125 \, ,
\end{equation}
and 
\begin{equation}
 h^{1,1}(B_3) = 11 \, ,\hspace{1cm} h^{2,1}(B_3)=0 \, ,
\end{equation}
giving the same spectrum by (\ref{eq:fthchiralmult}) and (\ref{eq:ftheoryvmult}).

\subsubsection{Dual Models with $G = SU(5)$}

Let us now consider an elliptic building block $Z_+$ with $N_+ = U \oplus A_4$, realizing $SU(5)$.  Such a model is found from the dual pair of tops with vertices
\begin{equation}
\begin{aligned}
 \Diamond^\circ &= \left(\begin{array}{rrrrrrrr}
-1 & 0 & 0 & 0 & 1 & 2 & 2 & 2 \\
0 & -1 & 0 & 1 & 1 & 3 & 3 & 3 \\
0 & 0 & 1 & 1 & 1 & 1 & -1 & 0 \\
0 & 0 & 0 & 0 & 0 & 0 & 0 & 1
\end{array}\right)  \\
\Diamond &= \left(\begin{array}{rrrrrrrrrrrr}
-2 & -1 & 1 & -1 & 1 & 1 & 1 & 1 & 1
& 1 & 0 & 0 \\
1 & 1 & 1 & 1 & 1 & 1 & 1 & 0 & 0 &
-1 & 0 & 0 \\
0 & -1 & 6 & -1 & 6 & -1 & -1 & -1 & -1
& 0 & -1 & -1 \\
0 & -2 & 0 & 0 & -6 & 0 & -6 & 0 & -3
& 0 & 0 & -1
\end{array}\right)\, .
\end{aligned}
\end{equation}
The corresponding hypersurface is determined as before and is a resolved $I_5$ Tate model. The matter loci are again points in the $Z_+$ building block, and an equal number of elliptic curves $\widehat{\mathbb{E}}$ on the F-theory side. 
The comparison of the spectra follows 
from \eqref{eq:bbtopology}, whereby
\begin{equation}
h^{1,1}(Z_+) =  7 \hspace{1cm} h^{2,1}(Z_+) =  76 \hspace{1cm} |N(Z_+)| = 6 \hspace{1cm} |K(Z_+)| = 0 \, ,
\end{equation}
so that
\begin{equation}
 b_2 = 16 \, , \hspace{1cm} b_3 = 223 \, .
\end{equation}

The dual F-theory geometry has 
\begin{equation}
h^{1,1}(X_F)=16 \, ,\hspace{1cm} h^{2,1}(X_F)=76 \, ,\hspace{1cm}  h^{3,1}(X_F)=100 \, ,
\end{equation}
as well as
\begin{equation}
 h^{1,1}(B_3) = 11 \, ,\hspace{1cm} h^{2,1}(B_3)=0 \,,
\end{equation}
again finding agreement. 

\subsubsection{Dual Models with $G = G_2$}

Non-simply laced gauge groups are found in situations where $G_{\rm root}\neq 0$, but the individual $\P^1$s of the resolution do not all become independent divisors on $Z_+$ and $X_F$. We can engineer a model with gauge symmetry $G_2$ by an appropriate fibration of a K3 surface with a single Kodaira $I_0^*$ fiber over $\P^1$ as $Z_+$. 
This fibration is of non-split type, in the language of \cite{Bershadsky:1996nh, Katz:2011qp}, i.e. there is monodromy acting on some of the resolution divisors in the K3-surface, such that they form one irreducible divisor in $Z_+$ and $X_F$. 
 Such a geometry can be captured by a dual pair of tops \cite{Candelas:1996su} with vertices 
\begin{equation}
 \begin{aligned}
 \Diamond^\circ = \left(\begin{array}{rrrrr}
-1 & 0 & 2 & 2 & 2 \\
0 & -1 & 3 & 3 & 3 \\
0 & 0 & 2 & -1 & 0 \\
0 & 0 & 0 & 0 & 1
\end{array}\right)\, ,\hspace{1cm}
\Diamond = \left(\begin{array}{rrrrrr}
-2 & 1 & 1 & 1 & 1 & 1 \\
1 & 1 & 1 & 1 & 1 & -1 \\
0 & 6 & 6 & -3 & -3 & 0 \\
0 & 0 & -6 & 0 & -6 & 0
\end{array}\right)\, .
 \end{aligned}
\end{equation}
From \eqref{eq:bbtopology} it follows that
\begin{equation}
h^{1,1}(Z_+) =  5 \hspace{1cm} h^{2,1}(Z_+) =  90 \hspace{1cm} |N(Z_+)| = 4 \hspace{1cm} |K(Z_+)| = 0 \, ,
\end{equation}
so that
\begin{equation}
 b_2 = 14 \, , \hspace{1cm} b_3 = 253 \, .
\end{equation}

The dual F-theory geometry has 
\begin{equation}
h^{1,1}(X_F)=14 \, ,\hspace{1cm} h^{2,1}(X_F)=90 \, ,\hspace{1cm}  h^{3,1}(X_F)=116 \, ,
\end{equation}
as well as
\begin{equation}
 h^{1,1}(B_3) = 11 \, ,\hspace{1cm} h^{2,1}(B_3)=0 \, .
\end{equation}

\subsubsection{Dual Models with a Gauge Group of Large Rank}

A particularly interesting building block $Z_+$ can be found from the largest projecting top, 
\begin{equation}
 \Diamond^\circ = \left(\begin{array}{rrrrr}
-1 & 0 & 2 & 2 & 2 \\
0 & -1 & 3 & 3 & 3 \\
0 & 0 & -1 & 6 & 6 \\
0 & 0 & 0 & 0 & 42
\end{array}\right)\,,
\end{equation}
which allows us to construct a model with a large rank $K(Z_+)$. 
It has
\begin{equation}
h^{1,1}(Z_+) =  251 \hspace{1cm} h^{2,1}(Z_+) =  0 \hspace{1cm} |N(Z_+)| = 10 \hspace{1cm} |K(Z_+)| = 240 \, , 
\end{equation}
so that matching with $Z_-$ gives 
\begin{equation}
b_2 = 260 \, ,\hspace{1cm} b_3 = 307 \, .
\end{equation}
This model has gauge group $E_8\times U(1)^{252}$. The dual F-theory geometry has 
\begin{equation}
 h^{1,1}(X_F)=260 \, ,\hspace{1cm} h^{2,1}(X_F)=240 \, ,\hspace{1cm}  h^{3,1}(X_F)=20 \, ,
\end{equation}
and
\begin{equation}
 h^{1,1}(B_3) = 251 \, ,\hspace{1cm} h^{2,1}(B_3)= 240 \, .
\end{equation}
It can be found from a nef partition of a polytope $\Delta^\circ$ with vertices
\begin{equation}
 \Delta = \left(\begin{array}{rrrrrrrrr}
-1 & 0 & 0 & 0 & 0 & 0 & 2 & 2 & 2 \\
0 & -1 & 0 & 0 & 0 & 0 & 3 & 3 & 3 \\
0 & 0 & 0 & 0 & 0 & 0 & -1 & 6 & 6 \\
0 & 0 & -1 & 0 & 0 & 0 & 0 & 0 & 42 \\
0 & 0 & 2 & -1 & 0 & 2 & 0 & 0 & 0 \\
0 & 0 & 3 & 0 & -1 & 3 & 0 & 0 & 0
\end{array}\right)
\end{equation}
This is an example with a highly non-trivial base $B_2$, where the $\mathbb{P}^1$-fiber degenerates into many reducible components, which is reflected in the large value for $|K(Z_+)|$.


\section{Extensions and Outlook}
\label{sec:Disc}

The main goal of this paper was to initiate the M-theory/Heterotic duality in the context of $G_2$ compactifications based on TCS constructions, and extending these to models with non-abelian gauge symmetry. These $G_2$-manifolds enjoy a K3-fibration and therefore seem to be particularly amenable for the study of this duality. To explicitly realize the M-theory/Heterotic duality fiberwise, we require the K3-fibers of the $G_2$-building blocks to be elliptically fibered. For smooth TCS $G_2$-manifolds, we identify the dual SYZ-fibered Calabi-Yau three-fold as the Schoen manifold on the heterotic side. The key generalization however is to singular K3-fibered TCS geometries, which we motivated both from heterotic and F-theory duals. The resulting models are 4d $N=1$ supersymmetric gauge theories with non-abelian gauge groups and non-chiral matter. These gauge groups can be of Higgsable or non-Higgsable type, and we provided constructions in both instances. 
There are numerous ways to utilize this setup for extensions and generalizations. 

The first question concerns a more careful analysis of the duality to heterotic, and a spectral cover description of the models including a spectral line bundle. Determining this, would then yield a first principle derivation via heterotic/F-theory duality of the absence of flux, which we infered in this paper by the match of the spectra. 
Futhermore we argued that due to the integrality of $c_2/2$ of the F-theory compactification, it is also consistent to not switch on $G_4$-flux.

Looking ahead, a natural question is to utilize the M-theory/heterotic duality also to include  conical singularities \cite{Acharya:2001gy}, which yield chiral matter in 4d. We were not able so far to identify these codimension seven singularities within the TCS construction, where singularities are by construction codimension six. 
The setup presented in this paper gives alternative ways to try to extend the TCS construction to include chirality. For instance, starting with the orbifold description of the Schoen manifold on the heterotic side, one could use magnetized tori instead, as in \cite{Nibbelink:2012de}. This generates chirality in the heterotic model, and understanding the dual in the M-theory construction may turn out to be insightful in guiding us towards realizing a  chiral spectrum from $G_2$ compactifications of M-theory. Likewise, chirality in F-theory is linked intimately with non-trivial $G_4$ flux, threading through matter surfaces. 

Finally, we should note, that the TCS construction  is suggestive of another limit, which is related to the work \cite{Pantev:2009de} on a Higgs bundle description, a local model given in terms of an ADE fibration over an associative cycle. We shall return to the connection of Higgs bundles and TCS geometries elsewhere. 


\subsection*{Acknowledgments}

We thank Jim Halverson, Alexei Kovalev, Magdalena Larfors,  Dave Morrison, Timo Weigand and Michele del Zotto, for discussions. 
We thank the Aspen Center for Physics for hospitality during the completion of this work, and the working group on `Physics and Geometry of $G_2$-manifolds' for discussions.  The Aspen Center for Physics is supported by National Science Foundation grant PHY-1607611. This work was also partially supported by Simons Foundation grant \# 488629.
AB would like to acknowledge support by the STFC grant ST/L000474/1.
SSN is supported by the ERC Consolidator Grant 682608 ``Higgs bundles: Supersymmetric Gauge Theories and Geometry (HIGGSBNDL)''.


\appendix

\section{Details of Geometric Constructions}

\subsection{Nef Partitions}\label{app:nef}

The combinatorial framework of \cite{Batyrev:1994hm} to construct families of Calabi-Yau hypersurfaces in toric varieties has an elegant extension to complete intersections in toric varieties \cite{Batyrev:1994pg}. As in the case of hypersurfaces, it starts with a pair of reflexive lattice polytopes which obey
\begin{equation}
 \langle \Delta,\Delta^\circ \rangle \geq -1 \, .
\end{equation}
The polytope $\Delta$ determines a toric variety via its normal fan, which can be refined using rays through lattice points on $\Delta^\circ$ to find an appropriate desingularization of the ambient space $A$ (which does not need to be a smooth space in general). A presentation of $\Delta^\circ$ as the convex hull of the sum of $n$ polytopes $\nabla_i$
\begin{equation}
\Delta^\circ = \langle \nabla_1,\cdots, \nabla_{n} \rangle_{\mbox{conv}}
\end{equation}
defines a nef partition if the dual polytopes $\Delta_i$ defined by
\begin{equation}
\langle \Delta_i , \nabla_j \rangle \geq -\delta_{ij}
\end{equation}
are lattice polytopes and recover $\Delta$ as their Minkowski sum
\begin{equation}
\Delta = \Delta_1 + \cdots + \Delta_{n} \, .  
\end{equation}
The corresponding nef partition is a decomposition 
\begin{equation}
[-K_{A}] = \sum_i [{\mathcal{L}_i}]
\end{equation}
into $n$ nef divisors classes. These determine the $n$ defining equations $P_i = 0$ of the complete intersection as
\begin{equation}
P_i = \sum_{m \in \Delta_i} c_m \prod_{\nu_k \in \Delta^\circ} z_k^{\langle m,\nu_k\rangle + \delta(\nu_k,\nabla_i)} = 0\, ,
\end{equation}
for a choice of coefficients $c_m$. Here, the sum runs over lattice points on $\Delta_i$ and the product runs over lattice points on $\Delta^\circ$. The function $\delta(\nu_k,\nabla_j)$ is one if the lattice point $\nu_k$ is on $\nabla_j$ and zero otherwise. 

The above framework allows a combinatorial computation of the Hodge numbers of the Calabi-Yau complete intersections. Swapping the roles of all of the $\Delta_i$ and the $\nabla_i$ realizes a construction of the mirror Calabi-Yau.

\subsection{The Schoen Calabi-Yau from a Nef Partition}\label{app:schoentoric}

In this section we give some details about how the realization of the Schoen Calabi-Yau $X_{\rm het}$ discussed in Section \ref{sect:schoentoric} as a Weierstrass double elliptic fibration over $\P^1$ is constructed as a toric complete intersection in the framework of Batyrev and Borisov. Consider the polytope
\begin{equation}
 \Delta^\circ = \left(\begin{array}{rrrrrrrr} 
-1 & 0 & 0 & 0 & 0 & 0 & 0 & 1 \\
0 & -1 & 0 & 0 & 0 & 0 & 2 & 2 \\
0 & 0 & -1 & 0 & 0 & 0 & 3 & 3 \\
0 & 0 & 0 & -1 & 0 & 2 & 0 & 2 \\
0 & 0 & 0 & 0 & -1 & 3 & 0 & 3
\end{array}\right)\, .
\end{equation}
The lattice points of $\Delta^\circ$ are the ray generators of a fan giving a toric variety with weight system
\begin{equation}
 \begin{array}{rrrrrrrr|c|rr}
\hat{z}_1 & \hat{x} & \hat{y} & x & y & w & \hat{w} & \hat{z}_2 & \Sigma \hbox{of degrees} & W&\hat{W} \\   
1 & 2 & 3 & 2 & 3 & 0 & 0 & 1 & 12 & 6 & 6 \\
0 & 2 & 3 & 0 & 0 & 0 & 1 & 0 & 6  & 0 & 6 \\
0 & 0 & 0 & 2 & 3 & 1 & 0 & 0 & 6 & 6 & 0
 \end{array}\, .
\end{equation}
One can confirm that there is a nef partition 
\begin{equation}
 \begin{aligned}[]
  [W] & = [x] + [y] + [w] + [\hat{z}_1] \\ 
  [\hat{W}]& =  [\hat{x}] + [\hat{y}] + [\hat{w}] + [\hat{z}_2] 
 \end{aligned}
\end{equation}
realizing \eqref{eq:eqschoentoric}. The vertices of the corresponding polytopes are 
\begin{equation}
\begin{aligned}
\nabla_W = \left(\begin{array}{rrrr}
0 & 0 & 0 & 1 \\
0 & 0 & 0 & 2 \\
0 & 0 & 0 & 3 \\
-1 & 0 & 2 & 2 \\
0 & -1 & 3 & 3
\end{array}\right) \hspace{1cm}
\nabla_{\hat{W}}=\left(\begin{array}{rrrr}
-1 & 0 & 0 & 0 \\
0 & -1 & 0 & 2 \\
0 & 0 & -1 & 3 \\
0 & 0 & 0 & 0 \\
0 & 0 & 0 & 0
\end{array}\right) \\ 
\Delta_W = 
\left(\begin{array}{rrrr}
-6 & 0 & 0 & 0 \\
0 & 0 & 0 & 0 \\
0 & 0 & 0 & 0 \\
1 & 1 & 1 & -2 \\
1 & 1 & -1 & 1
\end{array}\right)
\hspace{1cm}
\Delta_{\hat{W}} = \left(\begin{array}{rrrr}
-5 & 1 & 1 & 1 \\
1 & 1 & 1 & -2 \\
1 & 1 & -1 & 1 \\
0 & 0 & 0 & 0 \\
0 & 0 & 0 & 0
\end{array}\right)
\end{aligned}
\end{equation}
From this, it follows that the non-trivial Hodge numbers are $h^{1,1}(X_{\rm het})=h^{2,1}(X_{\rm het})=19$.

%
%

\subsection{$G_2$ Building Blocks}\label{app:bb}

In order to construct the building blocks $Z_{+,n}$, we start from the weight system \eqref{eq:weightsysfnbb}. Next, we construct a polyhedron such that the linear relations between its vertices give rise to the weights. We then neglect the coordinate $\hat{z}_2$ to find the associated top, which has the vertices
\begin{equation}
 \tilde{\Diamond}^\circ_{n}  = \left(\begin{array}{rrrrr}
-1 & 0 & 2 & 2 & 2 \\
0 & -1 & 3 & 3 & 3 \\
0 & 0 & -1 & n & 1 \\
0 & 0 & 0 & 1 & 0
\end{array}\right) \, .
\end{equation}
Its dual $\tilde{\Diamond}_{n}$, given by \eqref{eq:topsduality}, is not a projecting top for $n\geq 2$. The lattice points on $\tilde{\Diamond}_{n}$, however, still correspond to all of the monomials appearing in the defining equation $P_n=0$. What is missing is a resolution of singularities, which can be achieved by a subdivision of the fan, which can be thought of as  enlarging $\tilde{\Diamond}^\circ_n$. Correspondingly, the way $\tilde{\Diamond}_{n}$ fails to be a projecting top is by having vertices which are not lattice points. We can hence use the convex hull $\Diamond_{n}$ of all lattice points of $\tilde{\Diamond}_{n}$ instead and dualize back to find a polytope $\Diamond_{n}^\circ$ enlarging $\tilde{\Diamond}_{n}^\circ$. If this process generates a projecting top, it gives us a smooth manifold which is a crepant resolution of the singularities we started with\footnote{The same trick is of course applicable to Calabi-Yau hypersurfaces constructed from reflexive polytopes, and, as we will see in the next section, to Calabi-Yau complete intersections in toric varieties.}.

In the present case, this process generates a projecting top for each $n\leq 6$. For $n=5$, there is a locus in the base of its elliptic fibration over which $f,g,\Delta$ vanish with powers $4,6,12$, which we should blow up to have a clean effective field theory description. The blown up building block then corresponds to $\Diamond^{\circ *}_5$. This blow-up does not change the Hodge numbers of $Z_{+,5}$. 

The vertices of the polytopes we find are
\begin{equation}\label{eq:diamondsareforever}
\begin{aligned}
\Diamond^\circ_0 &= \left(\begin{array}{rrrrr}
-1 & 0 & 2 & 2 & 2 \\
0 & -1 & 3 & 3 & 3 \\
0 & 0 & -1 & 0 & 1 \\
0 & 0 & 0 & 1 & 0
\end{array}\right) \\
\Diamond^\circ_1 &= \left(\begin{array}{rrrrr}
-1 & 0 & 2 & 2 & 2 \\
0 & -1 & 3 & 3 & 3 \\
0 & 0 & -1 & 1 & 1 \\
0 & 0 & 0 & 0 & 1
\end{array}\right)\\
\Diamond^\circ_2 &= \left(\begin{array}{rrrrr}
-1 & 0 & 2 & 2 & 2 \\
0 & -1 & 3 & 3 & 3 \\
0 & 0 & -1 & 2 & 2 \\
0 & 0 & 0 & 0 & 1
\end{array}\right)\\ 
\Diamond^\circ_3 &= \left(\begin{array}{rrrrr}
-1 & 0 & 2 & 2 & 2 \\
0 & -1 & 3 & 3 & 3 \\
0 & 0 & -1 & 3 & 3 \\
0 & 0 & 0 & 0 & 1
\end{array}\right) \\
\Diamond^\circ_4 &=  \left(\begin{array}{rrrrrr}
-1 & 0 & 0 & 2 & 2 & 2 \\
0 & -1 & 1 & 3 & 3 & 3 \\
0 & 0 & 2 & -1 & 4 & 4 \\
0 & 0 & 0 & 0 & 0 & 1
\end{array}\right) \\
\Diamond^\circ_5 &= \left(\begin{array}{rrrrr}
-1 & 0 & 2 & 2 & 2  \\
0 & -1 & 3 & 3 & 3  \\
0 & 0 & -1 & 5 & 6  \\
0 & 0 & 0 & 1 & 0 
\end{array}\right)\\ 
\Diamond^{\circ *}_5 &= \left(\begin{array}{rrrrrr}
-1 & 0 & 2 & 2 & 2 & 2 \\
0 & -1 & 3 & 3 & 3 & 3 \\
0 & 0 & -1 & 5 & 6 & 6 \\
0 & 0 & 0 & 1 & 0 & 1
\end{array}\right)\\
\Diamond^\circ_6 &= \left(\begin{array}{rrrrr}
-1 & 0 & 2 & 2 & 2 \\
0 & -1 & 3 & 3 & 3 \\
0 & 0 & -1 & 6 & 6 \\
0 & 0 & 0 & 0 & 1
\end{array}\right)
 \end{aligned}
\end{equation}

\subsection{Calabi-Yau Fourfolds}\label{app:4folds}

To construct the Calabi-Yau fourfolds, we start from the weight system \eqref{eq:weightsftheory} and construct the associated fan with ray generators
\begin{equation}
 \begin{array}{rrrrrrrrrr}
 x & y & w & \hat{x} & \hat{y} & \hat{w} & z_1 & z_2 & \hat{z}_1 & \hat{z}_2 \\
 \hline
 0 & 0 & 0 & 0 & 0 & 0 & 0 & 0 & 1 & -1 \\
 0 & 0 & 0 & 0 & 0 & 0 & 1 &-1 & n & 0 \\
 0 & 0 & 0 & -1& 0 & 2 & 0 & 0 & 2 & 0 \\
 0 & 0 & 0 & 0 &-1 & 3 & 0 & 0 & 3 & 0 \\
 -1& 0 & 2 & 0 & 0 & 0 & 2 & 2 & 2 & 0 \\
 0 &-1 & 3 & 0 & 0 & 0 & 3 & 3 & 3 & 0 \\
 \end{array}\, .
\end{equation}
The convex hull $\tilde{\Delta}^\circ_n$ of these defines a reflexive polytope for $n\leq 1$ but fails otherwise. 

We can describe the resolved fourfolds $X_{Fn}$ for $n=0,\cdots,6$ by means of a nef partition as follows. The nef partition we are looking for must describe the pair of equations $W=\hat{W}=0$, i.e. 
\begin{equation}
\begin{aligned}
-K_A &= W + \hat{W} \\
W &= [\hat{z}_1] + [z_1] + [z_2] + [x] + [y] + [w] \\
\hat{W} &= [\hat{z}_2] + [\hat{x}] + [\hat{y}] + [\hat{w}] \,,
\end{aligned}
\end{equation}
which determines a decomposition of $\Delta^\circ$ into
\begin{equation}
 \tilde{\nabla}_W =  \left(\begin{array}{rrrrr}
0 & 0 & 0 & 1 & 0 \\
-1 & 0 & 0 & n & 1 \\
0 & 0 & 0 & 2 & 0 \\
0 & 0 & 0 & 3 & 0 \\
2 & -1 & 0 & 2 & 2 \\
3 & 0 & -1 & 3 & 3
\end{array}\right) \hspace{1cm}
\tilde{\nabla}_{\hat{W}} = \left(\begin{array}{rrrr}
-1 & 0 & 0 & 0 \\
0 & 0 & 0 & 0 \\
0 & -1 & 0 & 2 \\
0 & 0 & -1 & 3 \\
0 & 0 & 0 & 0 \\
0 & 0 & 0 & 0
\end{array}\right)\, .
\end{equation}
For $n\geq 2$, $\tilde{\Delta}^\circ$ is not reflexive, so that we are going to obtain a singular hypersurface. The polytopes of the dual nef partition, which are found from
\begin{equation}
\langle \tilde{\Delta}_i , \tilde{\nabla}_j \rangle \geq -\delta_{ij}
\end{equation}
are not lattice polytopes in this case. We can find a resolution by recalling that the $\tilde{\Delta}_i$ do contain all of the monomials appearing in the defining equations for $X_{Fn}$, but lack only a few blow-ups to cure the singularities enforced by the equations. We hence replace the $\tilde{\Delta}_i$ by the convex hull $\Delta_i$ of their integral points and dualize back to find polytopes $\nabla_i$ containing the $\tilde{\nabla}_i$ using the above duality relations. As $\Delta^\circ$ is the convex hull of the sum of the $\nabla_i$, this also enlarges $\tilde{\Delta}^\circ$ to $\Delta^\circ$, which corresponds to crepant (partial) resolutions of our singularities. If this results in a nef parition of a reflexive polytope $\Delta^\circ$, we have achieved our goal of resolving all singularities and can combinatorially compute the Hodge numbers.

In the present case, this method gives us nef partitions of reflexive polytopes for $n\leq 6$. In the case $n=5$, this gives a model with tensionless strings. We can blow up the base $B_3$ to excise such loci which takes us to a model which has an effective field theory description. We denote the corresponding polytope by $\Delta^{\circ *}_5$. The polytopes $\Delta_n^\circ$ are given by  

{ \footnotesize
\begin{equation}
\begin{aligned}
\Delta_2^\circ = \left(\begin{array}{rrrrrrrrr}
-1 & 0 & 0 & 0 & 0 & 0 & 0 & 0 & 1 \\
0 & -1 & 0 & 0 & 0 & 0 & 0 & 2 & 2 \\
0 & 0 & -1 & 0 & 0 & 0 & 2 & 0 & 2 \\
0 & 0 & 0 & -1 & 0 & 0 & 3 & 0 & 3 \\
0 & 2 & 0 & 0 & -1 & 0 & 0 & 2 & 2 \\
0 & 3 & 0 & 0 & 0 & -1 & 0 & 3 & 3
\end{array}\right) \\
\Delta_3^\circ = \left(\begin{array}{rrrrrrrrr}
-1 & 0 & 0 & 0 & 0 & 0 & 0 & 0 & 1 \\
0 & -1 & 0 & 0 & 0 & 0 & 0 & 3 & 3 \\
0 & 0 & -1 & 0 & 0 & 0 & 2 & 0 & 2 \\
0 & 0 & 0 & -1 & 0 & 0 & 3 & 0 & 3 \\
0 & 2 & 0 & 0 & -1 & 0 & 0 & 2 & 2 \\
0 & 3 & 0 & 0 & 0 & -1 & 0 & 3 & 3
\end{array}\right)\\
\Delta_4^\circ = \left(\begin{array}{rrrrrrrrrr}
-1 & 0 & 0 & 0 & 0 & 0 & 0 & 0 & 0 &
1 \\
0 & -1 & 0 & 0 & 0 & 0 & 0 & 2 & 4 &
4 \\
0 & 0 & -1 & 0 & 0 & 0 & 2 & 0 & 0 &
2 \\
0 & 0 & 0 & -1 & 0 & 0 & 3 & 0 & 0 &
3 \\
0 & 2 & 0 & 0 & -1 & 0 & 0 & 0 & 2 &
2 \\
0 & 3 & 0 & 0 & 0 & -1 & 0 & 1 & 3 &
3
\end{array}\right) \\
\Delta_5^{\circ}=\left(\begin{array}{rrrrrrrrr}
-1 & 0 & 0 & 0 & 0 & 0 & 0 & 0 & 1 \\
0 & -1 & 0 & 0 & 0 & 0 & 0 & 6 & 5 \\
0 & 0 & -1 & 0 & 0 & 0 & 2 & 0 & 2 \\
0 & 0 & 0 & -1 & 0 & 0 & 3 & 0 & 3 \\
0 & 2 & 0 & 0 & -1 & 0 & 0 & 2 & 2 \\
0 & 3 & 0 & 0 & 0 & -1 & 0 & 3 & 3
\end{array}\right)
\\
\Delta_5^{\circ *}= \left(\begin{array}{rrrrrrrrrr}
-1 & 0 & 0 & 0 & 0 & 0 & 0 & 0 & 1 & 1 \\
0 & -1 & 0 & 0 & 0 & 0 & 0 & 6 & 5 & 6 \\
0 & 0 & -1 & 0 & 0 & 0 & 2 & 0 & 2 & 2 \\
0 & 0 & 0 & -1 & 0 & 0 & 3 & 0 & 3 & 3 \\
0 & 2 & 0 & 0 & -1 & 0 & 0 & 2 & 2 & 2 \\
0 & 3 & 0 & 0 & 0 & -1 & 0 & 3 & 3 & 3
\end{array}\right) \\
\Delta_6^\circ = \left(\begin{array}{rrrrrrrrr}
-1 & 0 & 0 & 0 & 0 & 0 & 0 & 0 & 1 \\
0 & -1 & 0 & 0 & 0 & 0 & 0 & 6 & 6 \\
0 & 0 & -1 & 0 & 0 & 0 & 2 & 0 & 2 \\
0 & 0 & 0 & -1 & 0 & 0 & 3 & 0 & 3 \\
0 & 2 & 0 & 0 & -1 & 0 & 0 & 2 & 2 \\
0 & 3 & 0 & 0 & 0 & -1 & 0 & 3 & 3
\end{array}\right)
\end{aligned}
\end{equation}}



\providecommand{\href}[2]{#2}\begingroup\raggedright\endgroup

\end{document}